\documentclass[fleqn,usenatbib]{mnras}

\usepackage{newtxtext,newtxmath}

\usepackage[T1]{fontenc}

\DeclareRobustCommand{\VAN}[3]{#2}
\let\VANthebibliography\thebibliography
\def\thebibliography{\DeclareRobustCommand{\VAN}[3]{##3}\VANthebibliography}

\usepackage{graphicx}	
\usepackage{amsmath}	

\title[Gargantuan chaotic triple systems]{Gargantuan chaotic gravitational three-body systems II. Dependence on angular momentum and astrophysical scale}

\author[T. Boekholt and S. Portegies Zwart]{
T. C. N. Boekholt$^{1}$\thanks{E-mail: tjardaboekholt@gmail.com (TB)}
and S. F. Portegies Zwart,$^{2}$\thanks{E-mail: spz@strw.leidenuniv.nl (SPZ)}
\\
$^{1}$Rudolf Peierls Centre for Theoretical Physics, Clarendon Laboratory, University of Oxford, Parks Road, Oxford, OX1 3PU, UK\\
$^{2}$Leiden Observatory, Leiden University, PO Box 9513, 2300 RA, Leiden, The Netherlands
}

\date{Accepted XXX. Received YYY; in original form ZZZ}

\pubyear{2015}

\begin{document}
\label{firstpage}
\pagerange{\pageref{firstpage}--\pageref{lastpage}}
\maketitle

\begin{abstract}
Recently we estimated that about 5 percent of supermassive black hole triple systems are fundamentally unpredictable. These gargantuan chaotic systems are able to exponentially magnify Planck length perturbations to astronomical scales within their interaction timescale. These results were obtained in the zero angular momentum limit, which we naively expected to be the most chaotic. Here, we generalise to triple systems with arbitrary angular momenta by systematically varying the initial virial ratio. We find the surprising result that increasing the angular momentum enhances the chaotic properties of triples. This is not only explained by the longer life times, allowing for a prolonged exponential growth, but also the maximum Lyapunov exponent itself increases. For the ensemble of initially virialised triple systems, we conclude that the percentage of unpredictable supermassive black hole triples increases to about 30 percent. A further increase up to about 50 percent is reached when considering triples on smaller astrophysical scales. Fundamental unpredictability is thus a generic feature of chaotic, self-gravitating triple populations. 
\end{abstract}




\section{Introduction}

The gravitational three-body problem \citep{Newton:1687} is ubiquitous
in the Universe. Examples range from the Sun--Earth--Moon system \citep[e.g.][]{Touma_1998}, to
Trans-Neptunian triplets \citep[e.g.][]{2018Icar..305..250C}, Jacobi captures \citep[e.g.][]{TB23}, hierarchical
triple stars \citep[e.g.][]{Toonen2022}, stellar-mass black holes \citep[e.g.][]{2000ApJ...528L..17P,2014ApJ...784...71S}, supermassive black holes
\citep[e.g.][]{2020A&A...633A..79K, TB2021}, and even whole galaxies
\citep{2008MNRAS.384..886V}.

Chaos in the gravitational three-body problem was first demonstrated
by Henri Poincar\'e \citep{Poincare91, Poincare92}. It manifests
itself through the exponential sensitivity to small perturbations
\citep{1964ApJ...140..250M, 1986LNP...267..212D,
  1993ApJ...415..715G}. This exponential growth of a small disturbance in the force
is thought to be driven by the non-linear influence of the third
body. If the finite-mass third body is infinitely far away, the
motions of the other two bodies can be obtained analytically by
solving the Kepler-problem \citep{1609asno.book.....K}. If the third body is systematically brought
closer to the other two bodies, its gravitational influence increases
and the gravitational interaction becomes non-linear; this leads to
chaos. Here we define chaotic response as the exponential growth of
any small perturbation in the motion of the three bodies, and regular response when the growth is weaker than exponential.

The motion in non-hierarchical three-body systems is characterised by
a series of transitions between regular and chaotic motion.  This
behaviour manifests itself in sequences of short democratic resonances
with extended phases of hierarchical evolution \citep{MH96}.  This latter phase can be represented by L\'evy flights \citep{2019Natur.576..406S,
  2020MNRAS.497.3694M} in phase space. However, not all variations and transitions in the exponential growth rate can be explained this way, indicating the presence of other mechanisms also affecting the rate of divergence
\citep{2018CNSNS..61..160P, SPZ23}.

To mediate the discussion we will quantify chaotic motion by a
measure of the local Lyapunov time scale $T_{\rm \lambda}$
\citep{2018CNSNS..61..160P}. This measure is the reciproke of the local, finite-time, maximum Lyapunov exponent of a system. The value of this chaotic quantifier is local (in time), and although its value can vary considerably from one moment to the next, the global trend is surprisingly stable. This leads to well-defined averaged Lyapunov exponents, which are an inherent property of chaotic N-body systems \citep{TB2020, SPZ23}. The relation between the Lyapunov exponent of a system and the detailed motions of the three bodies is still an open problem. Lines of investigation include orbital resonances and resonance overlap, or sequences of strong two-body encounters and deflections
\citep{1993ApJ...415..715G, 2016MNRAS.461.3576B}, i.e. punctuated chaos \citep{SPZ23}. 

Irrespective of the mechanism that drives the exponential divergence between two neighbouring trajectories in phase space, exponential sensitivity can be measured through performing accurate and precise N-body simulations \citep{TB2020}.  Here the accuracy reflects a measure to which energy is conserved, and precision can be perceived as the number of decimal places used to express the result \citep{2018CNSNS..61..160P}.
\cite{TB2020} (hereafter Paper 1) studied triples systematically drawn from the Agekyan-Anosova
map \citep{1967AZh....44.1261A,1968SvA....11.1006A}, and measured a power law distribution of
amplification factors, e.g. the amplification factor of a small
initial perturbation over the lifetime of the triple. The implication is the presence of a power law tail of triple systems, which produce extremely large amplification factors, i.e. gargantuan chaotic triple systems.
We found a finite fraction of systems that would remain irreversible, even if the calculations would be conducted with a precision below the Planck length. We therefore argued that these gargantuan chaotic systems would be fundamentally irreversible and unpredictable. Despite our lack of
understanding on the deterministic behaviour of quantum gravity at
sub-Planck length scales and considering Heisenberg's principle of
uncertainty, we argue that this irreversible fraction may be
fundamental to the arrow of time \citep{2022arXiv220903347P}.

In Paper 1 we focused on the amplification
factor for zero-angular momentum orbits, naively assuming that those
would be more chaotic than configurations with non-zero angular
momentum.  
Our naivety in the zero-angular momentum problem stems from the fact
that for two-body problems, smaller angular momentum implies more
radial orbits. As a consequence, and extrapolating to three-body
systems, this naturally leads to close encounters. On the other hand,
bodies in configurations with finite angular momentum would remain on average at larger distance from each other, leading to fewer and weaker
interactions. Based on such an intuition it might be argued that zero-angular momentum systems would be maximally chaotic. As it turns out now, as we explain in this paper, this only holds statistically for the shortest lived triple systems (up to about 20 crossing times). We find the surprising result that the opposite is true for longer lived systems (which are also the majority); more angular momentum in three-body initial conditions leads to longer lifetimes, shorter Lyapunov time scales, larger magnification factors, and therefore to more chaos.  Generally, when increasing the initial amount of angular momentum, the fraction of irreversible solutions increases. 

This second paper on gargantuan chaotic triples extends the results of Paper 1 to triples with arbitrary angular momenta, and demonstrates that the fraction of unpredictable triples can reach larger values. We will also scale the results to triples of arbitrary physical scale finding that the fraction of unpredictable triples is further enhanced for more compact triples, such as triple asteroids (see Sec.~\ref{sec:scale}). Complementary to the intrinsic quantum perturbations, we will also briefly discuss tidal perturbations from the presence of other bodies in the Universe. In most cases these tend to dominate over intrinsic Planck length uncertainties, which enhances the unpredictability of N-body systems even further (see Sec.~\ref{sec:isolation}).  

\section{Methods}

\subsection{Initial conditions}\label{sec:setup}

We adopt the equal mass Plummer distribution \citep{Plummer_1911} for drawing random initial positions and velocities for the three bodies, resulting in an unstable triple system configuration (similar to e.g. \citet{2015ComAC...2....2B}). The center of mass is set to be at rest on the origin, and the coordinates are rescaled to H\'enon units \citep[e.g.][]{1986LNP...267..233H} for which

\begin{equation}
G = 1,
\end{equation}
\begin{equation}
M = 1,
\end{equation}
\begin{equation}
E = -\frac{1}{4},   
\end{equation}

\noindent where $G$ is the gravitational constant, $M$ is the total mass, and $E$ the total energy. The virial radius serves as a characteristic size of the triple and is given by 

\begin{equation}
R_v = -\frac{G M^2}{4E} = 1. 
\end{equation}

\noindent Similarly, the characteristic speed of a body, $\sigma$, is given by

\begin{equation}
    \sigma = \sqrt{-\frac{2E}{M}} = \frac{1}{\sqrt{2}}.
\end{equation}

\noindent The characteristic dynamical time (or crossing time) is then defined as

\begin{equation}
    T_c = \frac{2 R_v}{\sigma} = 2 \sqrt{2}.
\end{equation}

So far, the angular momentum, $L$, of the triple does not appear. Using the above prescription to generate triple realisations, we indeed find that $L$ is not constant among the triples. The average value for the virialised case is $\langle L_{\rm{virial}} \rangle = 0.19$.

In order to generate ensembles of triples with a lower average angular momentum, we adjust the initial virial ratio, $Q$, which is defined as the ratio between total kinetic energy, $T$, and absolute value of the potential energy, $V$: 

\begin{equation}
    Q = \frac{T}{|V|}.
\end{equation}

\noindent A triple is best served cold, i.e. with zero velocities, and has a virial ratio of $Q = 0$ and angular momentum $L = 0$. By
systematically increasing the initial velocities, we increase both $Q$ and $L$, until we eventually reach the initially virialised configuration.

The procedure for generating initial conditions starts by generating a random realisation of a virialised triple system drawn from a Plummer distribution. For the virial case we have the relation $2T + V = 0$, and therefore $Q = \frac{1}{2}$. To change the virial ratio we rescale the velocities, and therefore $T$, by a factor $C_V$:

\begin{equation}
    \frac{Q}{Q_V} = \frac{T}{T_V} = \left( \frac{\sigma}{\sigma_V} \right)^2 \equiv C_v^2,
\end{equation}

\noindent with $\sigma$ the mass-weighted velocity dispersion, and the subscript $V$ refers to virial. Since $Q_v = \frac{1}{2}$, we can write

\begin{equation}
C_v = \sqrt{2 Q}.
\end{equation}

\noindent This scaling of the velocities changes the total energy of the triple to a value 

\begin{equation}
    E_1 = \frac{1}{2} M \left( C_v \sigma_V \right)^2 - \frac{1}{2}\frac{G M^2}{R_V} = \frac{1}{4} C_v^2 - \frac{1}{2} = \frac{Q}{2} - \frac{1}{2},
\end{equation}

\noindent whereas its initial virialised value was set to $E_0 = -\frac{1}{4}$. In order to normalise the energy, we calculate the scaling factor of the total energy

\begin{equation}
C_E \equiv \frac{E_0}{E_1} = \frac{1}{2 - C_V^2} = \frac{1}{2-2Q},
\end{equation}

\noindent and rescale the positions and velocities by factors respectively given by

\begin{equation}
D_r \equiv \frac{1}{C_E} = 2-C_V^2 = 2-2Q,    
\end{equation}
\begin{equation}
D_v \equiv \sqrt{C_E} = \frac{1}{\sqrt{2-C_V^2}} = \frac{1}{\sqrt{2-2Q}},
\end{equation}

\noindent which rescales the total energy back to a value $E_0$, while preserving the virial ratio $Q$. The total angular momentum scales as

\begin{equation}
    \frac{L}{L_V} = D_r C_v D_v = \frac{\sqrt{2 Q}\left( 2-2Q \right)}{\sqrt{2-2Q}} = 2 \sqrt{Q \left( 1 - Q \right) }, 
    \label{eq:virial}
\end{equation}

\begin{figure}
\centering
\begin{tabular}{c}
\includegraphics[height=0.384\textwidth,width=0.48\textwidth]{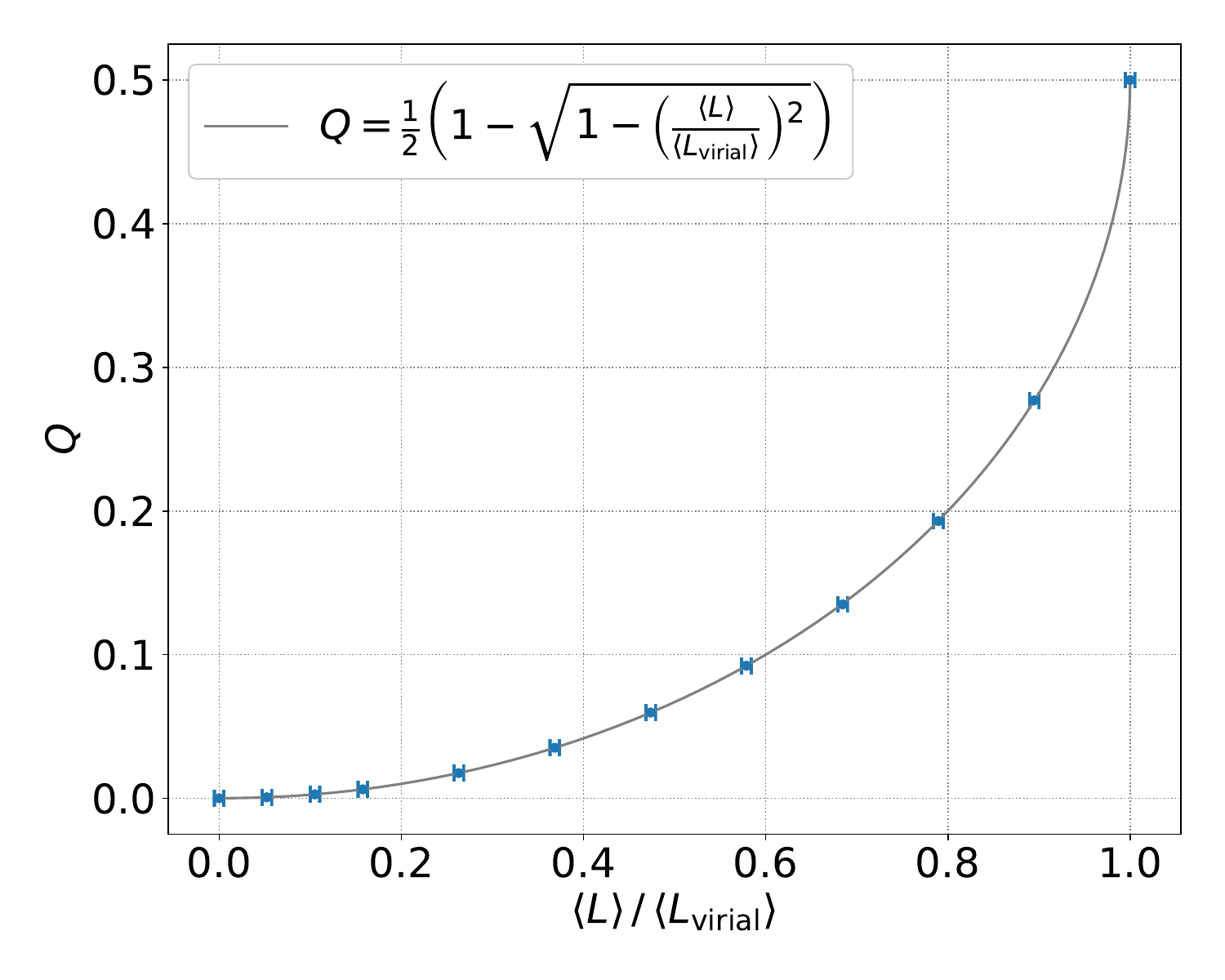} \\
\end{tabular}  
\caption{ Relation between virial ratio, $Q$, and angular momentum, $L$. Our ensembles of triple systems vary $L$ systematically and are confined to be within the horizontal ``error bars''. }
\label{fig:ics}
\end{figure}

\noindent where $D_r$ is the net scaling of the positions, and $C_v D_v$ is the net scaling of the velocities. From this expression we confirm that for $Q=0$ we obtain $L=0$, while for $Q=\frac{1}{2}$ we retrieve $L=L_V$. The drawing of random realisations of triple systems for a given value of $Q$ still results in a spread of values for $L$. The average value of $L$ however does scale one-to-one with $Q$, and we therefore use

\begin{equation}
    \frac{\langle L\rangle}{\langle L_{\rm{virial}} \rangle} = 2 \sqrt{Q \left( 1-Q \right)},
\end{equation}

\noindent or after inverting the equation we obtain 

\begin{equation}
    Q = \frac{1}{2} \left( 1 - \sqrt{ 1 - \left( \frac{\langle
        L\rangle}{\langle L_{\rm{virial}}\rangle }  \right)^2 } \right).
\end{equation}

\noindent Henceforth, we might omit the angular brackets ($L$ instead of $\langle L \rangle$) for simplicity, i.e. $L$ refers to the average value of the ensemble unless stated otherwise. 
In Fig.~\ref{fig:ics} we plot the relation between $Q$ and $L$. We define 12 ensembles of triple systems for which the average angular momentum is systematically varied. For each ensemble, we calculate the corresponding value for $Q$. Then we generate 2,048 random, virialised triple realisations, which we subsequently rescale according to the procedure described above. Since we are interested in measuring trends with $L$, we do not want the range in $L$ within an ensemble to be too broad. Therefore, we only accept a random realisation of a triple if its angular momentum is within the range given by $L \pm \delta L$ with $\delta L = 0.001$. For the virialised case, this margin corresponds to $\delta L / L_{\rm{virial}} = 0.0053$. 
The corresponding values of $L$ and $Q$ are given in Tab.~\ref{tab:ics}, and the ensembles are also presented in Fig.~\ref{fig:ics} as the horizontal ``errorbars''. It is clear that the various ensembles do not overlap, which allows us to measure statistical trends as a function of angular momentum. 

\begin{table*}
\centering
\begin{tabular}{| lll | cccccc | cc | cc | } 
\hline
$\langle L \rangle / \langle L_{\rm{virial}} \rangle$ & $\langle L \rangle$ & Q & $\nu$ & $\delta \nu$ & $\eta$ & $\delta \eta$ & $\alpha$ & $\delta \alpha$ & $\beta$ & $\delta \beta$ & $p\,\rm{[\%]}$ & $\delta p\,\rm{[\%]}$\\
 \hline\hline
1.00   & 0.19  & 0.5      & 
0.0115  & 0.0003 & 0.060  & 0.003 & 0.83 & 0.01 & 0.12 & 0.02 & 29.8  & 0.4 \\ 
0.895  & 0.17  & 0.277    & 
0.0131  & 0.0004 & 0.074  & 0.003 & 0.88 & 0.01 & 0.06 & 0.02 & 25.4  & 0.5 \\ 
0.790  & 0.15  & 0.193    & 
0.0147  & 0.0002 & 0.079  & 0.002 & 0.86 & 0.02 & 0.08 & 0.03 & 21.5  & 0.2 \\ 
0.684  & 0.13  & 0.135    & 
0.0170  & 0.0001 & 0.089  & 0.001 & 0.82 & 0.02 & 0.15 & 0.02 & 16.6  & 0.1 \\ 
0.579  & 0.11  & 0.0923   & 
0.0206  & 0.0005 & 0.106  & 0.005 & 0.80 & 0.02 & 0.18 & 0.03 & 11.4  & 0.3 \\ 
0.474  & 0.09  & 0.0597   & 
0.0237  & 0.0003 & 0.125  & 0.002 & 0.76 & 0.01 & 0.23 & 0.02 & 8.2  & 0.1 \\ 
0.368  & 0.07  & 0.0352   & 
0.0270  & 0.0005 & 0.147  & 0.004 & 0.79 & 0.02 & 0.19 & 0.03 & 5.9  & 0.2 \\ 
0.263  & 0.05  & 0.0176   & 
0.030   & 0.001 & 0.163  & 0.009 & 0.71 & 0.01 & 0.29 & 0.02 & 4.2  & 0.2 \\ 
0.158  & 0.03  & 0.00627  & 
0.0322  & 0.0006 & 0.172  & 0.005 & 0.68 & 0.01 & 0.31 & 0.02 & 3.4  & 0.1 \\ 
0.105  & 0.02  & 0.00278  & 
0.033   & 0.001 & 0.18   & 0.01 & 0.75 & 0.02 & 0.17 & 0.02 & 3.1  & 0.2 \\ 
0.0524 & 0.01  & 0.000693 & 
0.0291  & 0.0007 & 0.154  & 0.005 & 0.66 & 0.02 & 0.34 & 0.03 & 4.7  & 0.2 \\ 
0.0    & 0.0   & 0.0      & 
0.0282  & 0.0006 & 0.151  & 0.005 & 0.75 & 0.01 & 0.19 & 0.02 & 5.2  & 0.2 \\ 
\hline
\end{tabular}
\caption{Table of initial conditions and statistical outcomes. The relation between virial ratio, $Q$, and angular momentum, $L$, is visualised in Fig.~\ref{fig:ics}. The parameters $\nu$ and $\eta$ correspond to the linear fits in Fig.~\ref{fig:firr}. The parameters $\alpha$ and $\beta$ correspond to the linear fits in Fig.~\ref{fig:logA_vs_T_gallery}. The parameter $p$ refers to the percentage of fundamentally unpredictable triples in our massive black hole application (see Sec.~\ref{sec:discussion}). }
\label{tab:ics}
\end{table*}

\subsection{Experimental setup}

Adopting the same strategy as \citet{2018CNSNS..61..160P} and \citet{TB2020}, we use the code \texttt{Brutus} \citep{2015ComAC...2....2B} to perform a reversibility test. Each initial condition is integrated forwards in time until dissolution of the triple configuration. We define this to be the case if one of the bodies 1) is at a distance beyond 10 H\'enon units from the center of mass of the triple, 2) is moving away from the center of mass, and 3) has a positive energy, meaning it has become unbound from the remaining binary system. Since a small fraction of triples is very long lived, we also set a maximum simulation time of $10^4$ H\'enon time units, or about $3536\,T_c$.
We define the duration of the forward integration from the initial condition to the stopping conditions as the lifetime, $T$, of the triple. At that moment, we flip the sign of each velocity coordinate of each particle, and integrate forwards until a final simulation time of $t = 2 T$. Note that effectively, the system is evolving back to the future initial condition. After flipping the velocity signs once more, we can compare the initial (forward) and final (backward) configurations by measuring their phase space distance, given by

\begin{equation}
\Delta^2 = \sum_{i=1}^{N}\,\sum_{k=1}^{6}\,\left( x_{i,k,f} - x_{i,k,b} \right)^2,     
\end{equation}

\noindent where the first sum is over all bodies ($N=3$), the second sum over all phase space coordinates (positions and velocities), and where the subscript $f$ and $b$ refer to the forward and backward integration, respectively. Or in words, the phase space distance, $\Delta$,  gives the Euclidean distance between the two solutions in $6N$-dimensional phase space. For a perfectly time-reversible integrator with reversible numerical errors, the phase space distance between the forward and backward solutions would remain zero. In this case, a reversibility test does not hold any information about the accuracy of the solution. However, in the presence of irreversible numerical errors, these errors will serve as a perturbation to the system, which seeds the exponential divergence between the forward and backward integrations, i.e. the phase space distance between the two solutions will grow exponentially. A reversibility test is then declared successful if the phase space distance between the initial and final states is below some small threshold, for which we adopt

\begin{equation}
    \log_{10}\,\Delta \leq -3. 
\end{equation}

\noindent This criterion ensures we remain in the linear perturbation regime, as $\Delta$ remains a factor $10^3$ smaller than the characteristic size and speed of the triple system.

In our first attempt to obtain a reversible solution for each triple, we set the Bulirsch-Stoer tolerance parameter to $\epsilon = 10^{-6}$, and we express the word-length (length of the mantissa in units of bits) as 

\begin{equation}
L_w = -4 \log_{10}\,\epsilon + 32,
\end{equation}

\noindent which for $\epsilon=10^{-6}$ corresponds to 56 bits, and more bits are added as $\epsilon$ decreases \citep{2015ComAC...2....2B}. If the reversibility test fails for a subset of triples, then we redo the test with a smaller value of $\epsilon$ and the corresponding value of $L_w$. This way, we can measure the fraction of irreversible solutions as a function of numerical accuracy and precision. We halted the iteration at $\epsilon = 10^{-90}$, but we will show that even then a fraction of triples still remained irreversible. Besides the fraction of irreversible solutions, we also measure the two main observables for each triple, which are the lifetime $T$, and the amplification factor defined as

\begin{equation}
    A = \frac{\Delta_f}{\Delta_i}.
\end{equation}

\noindent Here, $\Delta_i$ is the phase space distance between the forward and backward solution after a single integration step (i.e. the states just before and after $t=T$ are compared), and $\Delta_f$ is the phase space distance between the initial and final states. In other words, $A$ gives the total amplification factor of the initial perturbation over the lifetime of the triple. The finite-time Lyapunov exponent is then estimated as 

\begin{equation}
    \lambda = \frac{\log\,A}{T},
\end{equation}

\noindent while the finite-time Lyapunov time scale is the inverse, i.e. $T_\lambda = \lambda^{-1}$.

\begin{figure}
\centering
\begin{tabular}{c}
\includegraphics[height=0.384\textwidth,width=0.48\textwidth]{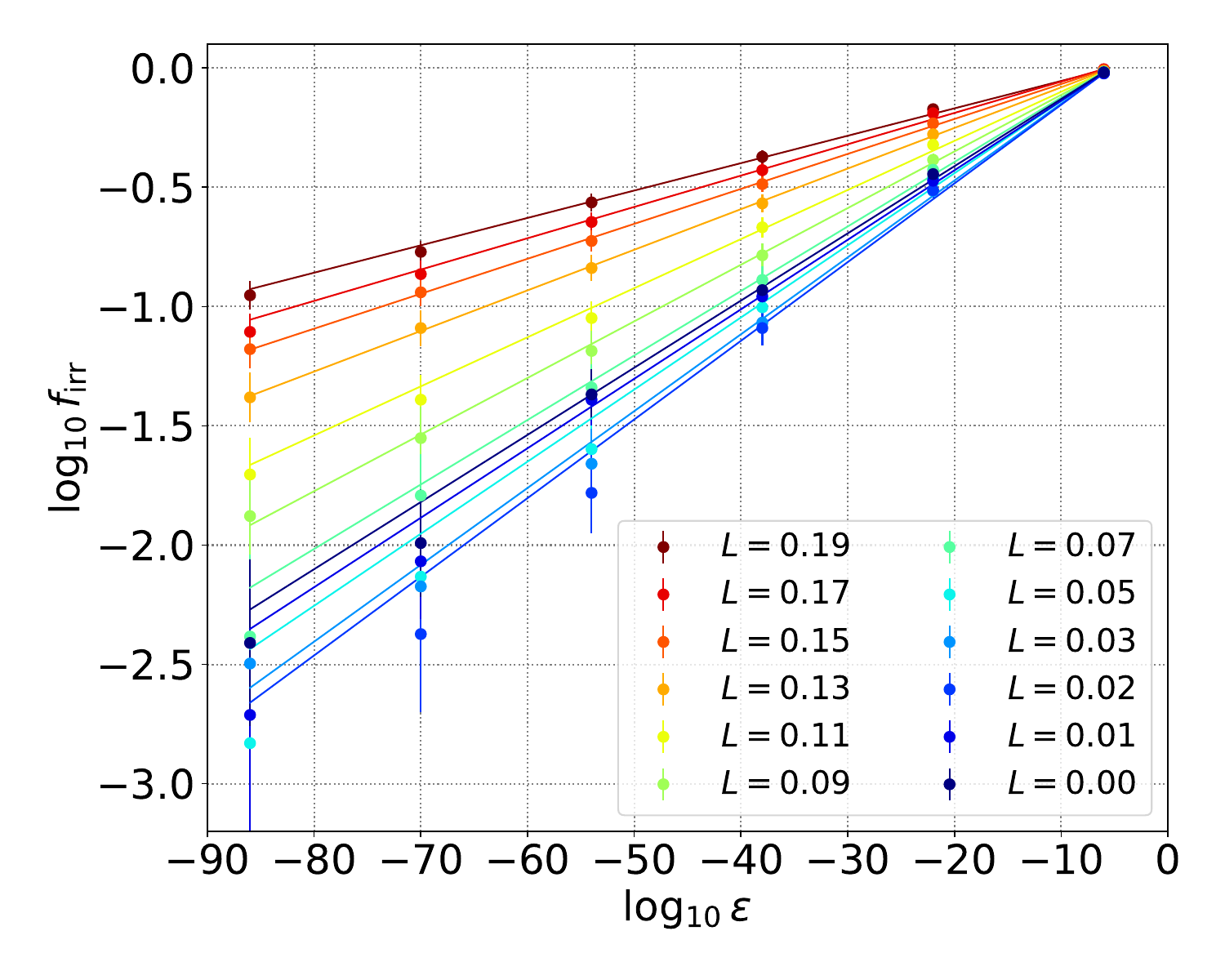} \\
\end{tabular}  
\caption{ Fraction of irreversible simulations, $f_{\rm{irr}}$, as a function of integration accuracy, which is represented by the Bulirsch-Stoer tolerance, $\epsilon$. The data points and errorbars are obtained from the data, while the curves are linear fits whose parameters are given in Tab.~\ref{tab:ics}, with $\mu$ the slope and $\eta$ the offset.
Higher angular momentum triples have a larger fraction of irreversible solutions. Interestingly, we observe that the lowest irreversible fraction is produced by $L=0.02$ rather than $L=0$. This turns out to be correlated with statistically shorter lifetimes (see Fig.~\ref{fig:cdfs}). }
\label{fig:firr}
\end{figure}

\begin{figure}
\centering
\begin{tabular}{c}
\includegraphics[height=0.24\textwidth,width=0.48\textwidth]{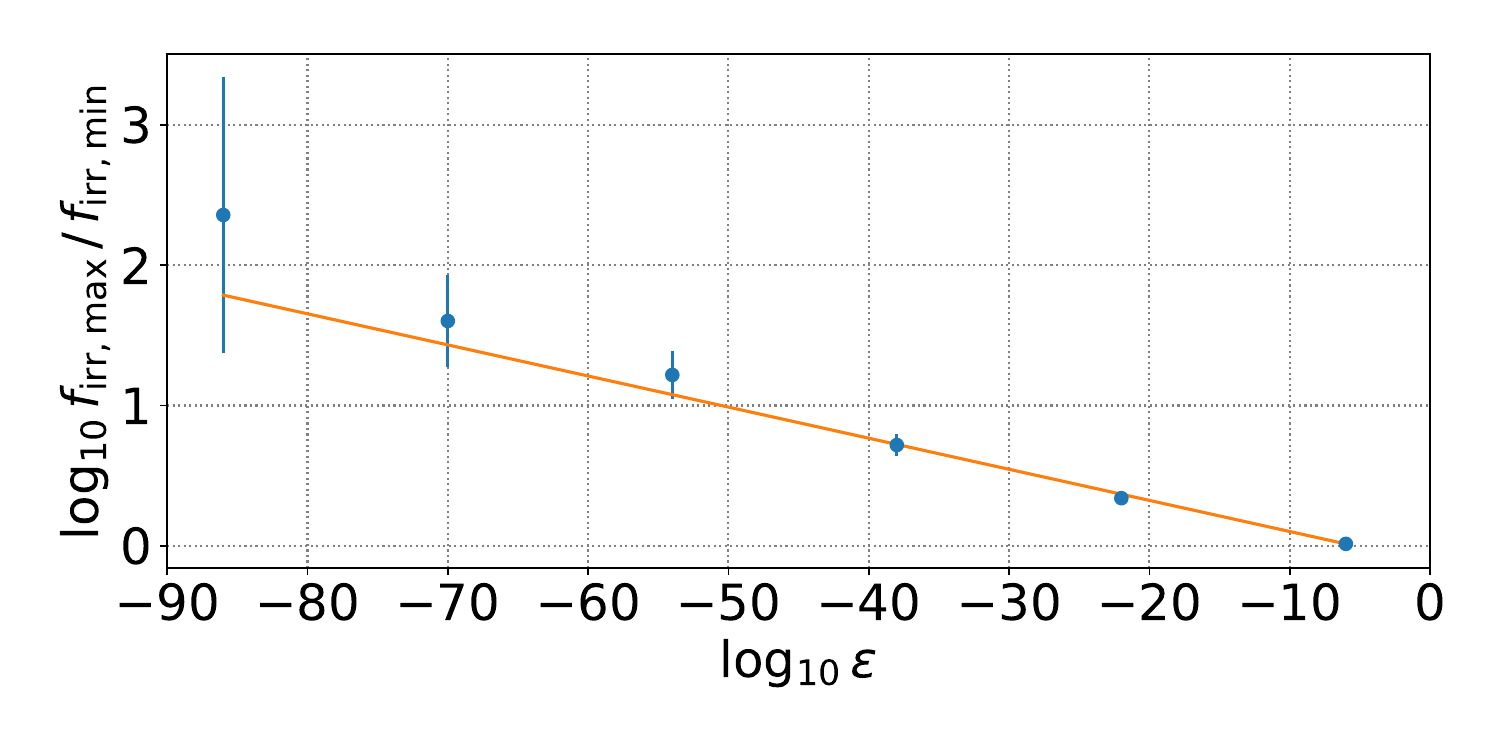} \\
\end{tabular}  
\caption{ We plot the ratio between the largest and smallest irreversible fraction from Fig.~\ref{fig:firr}, i.e. between those of $L=0.19$ and $L=0.02$. We fit a linear relation resuling in a slope $-0.022 \pm 0.001$ and offset $-0.120 \pm 0.008$.  }
\label{fig:firr_diff}
\end{figure}

\section{Results}

\begin{figure*}
\centering
\begin{tabular}{ccc}
\includegraphics[height=0.289\textwidth,width=0.32\textwidth]{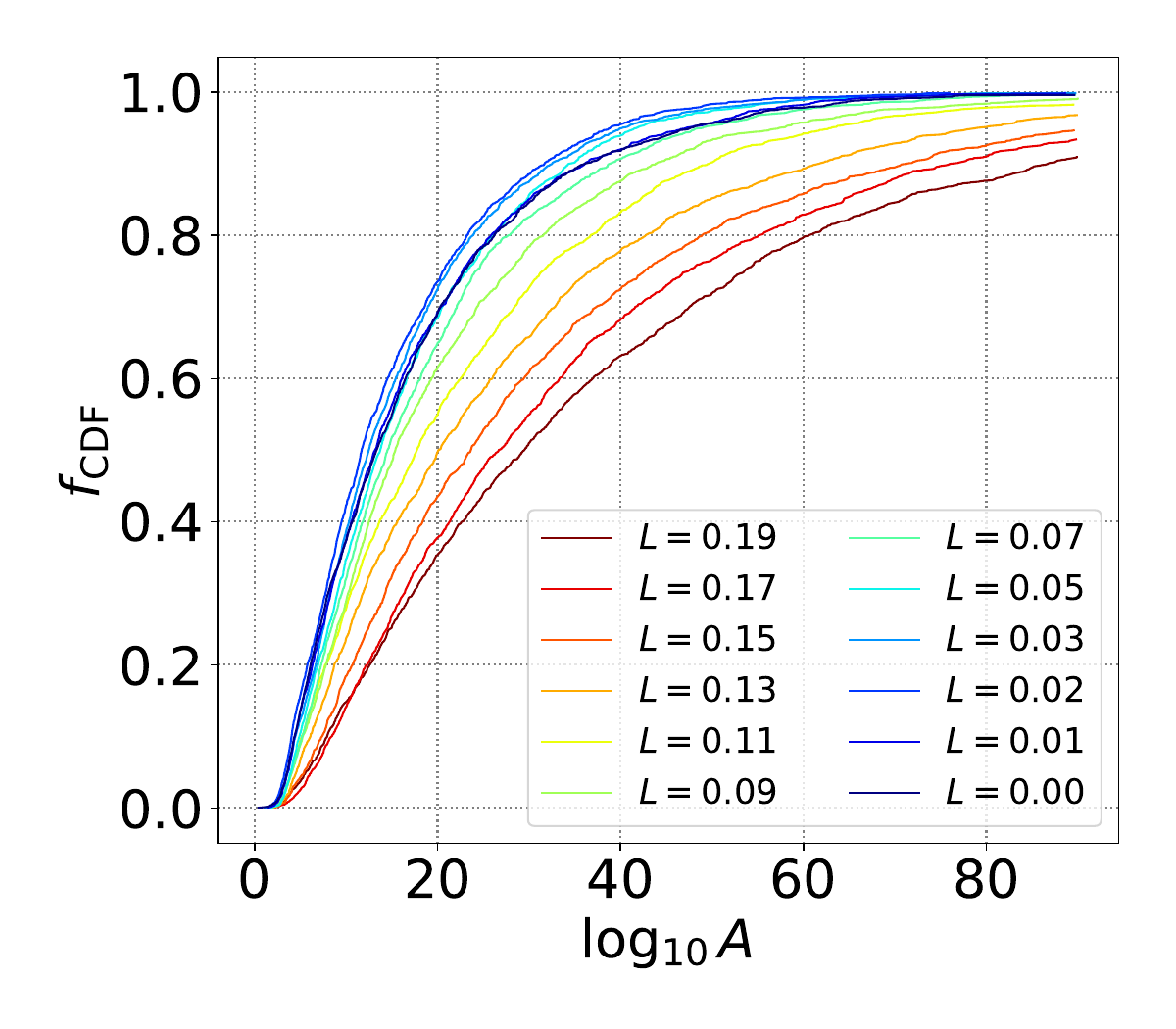} &
\includegraphics[height=0.289\textwidth,width=0.32\textwidth]{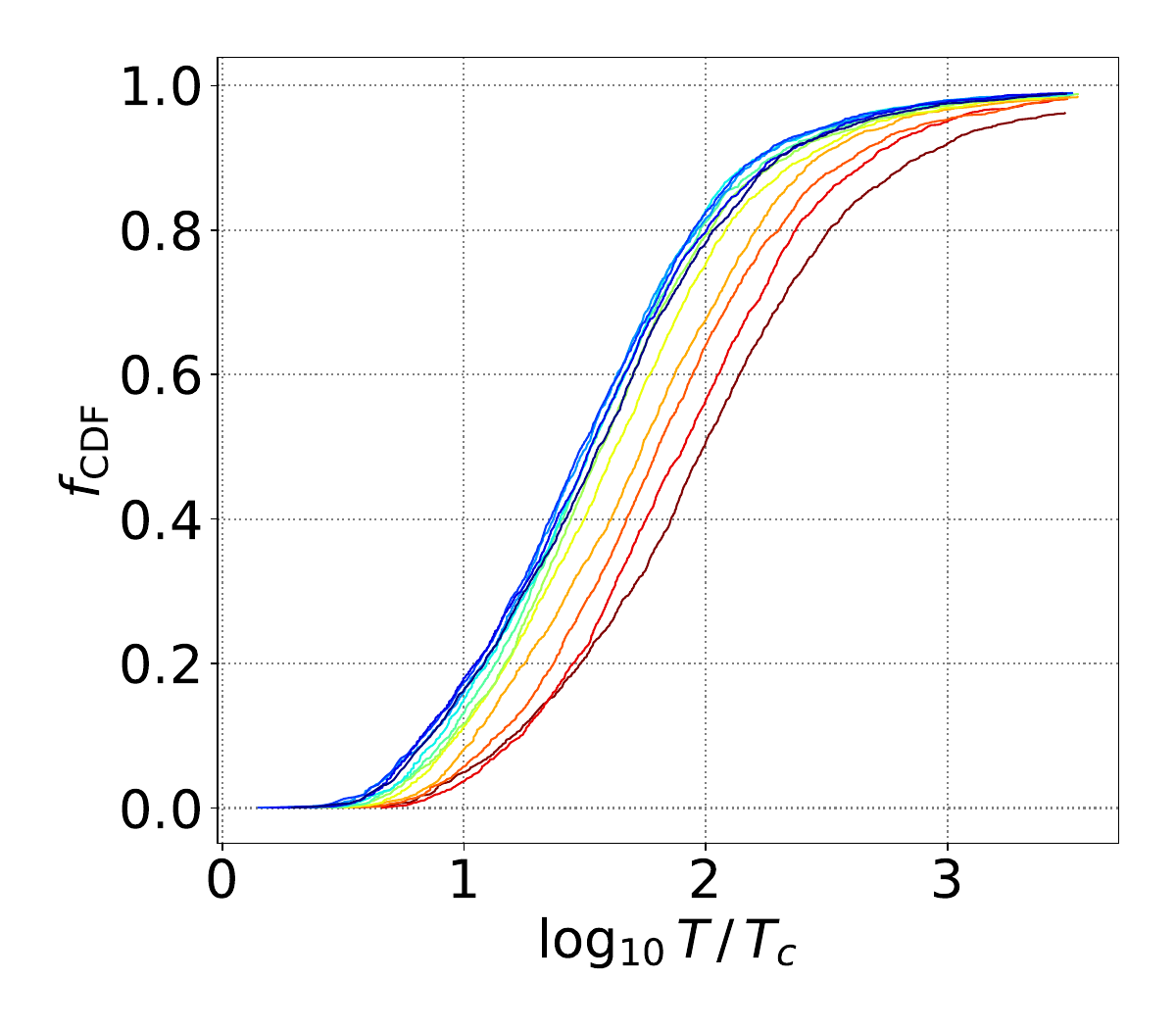} &
\includegraphics[height=0.289\textwidth,width=0.32\textwidth]{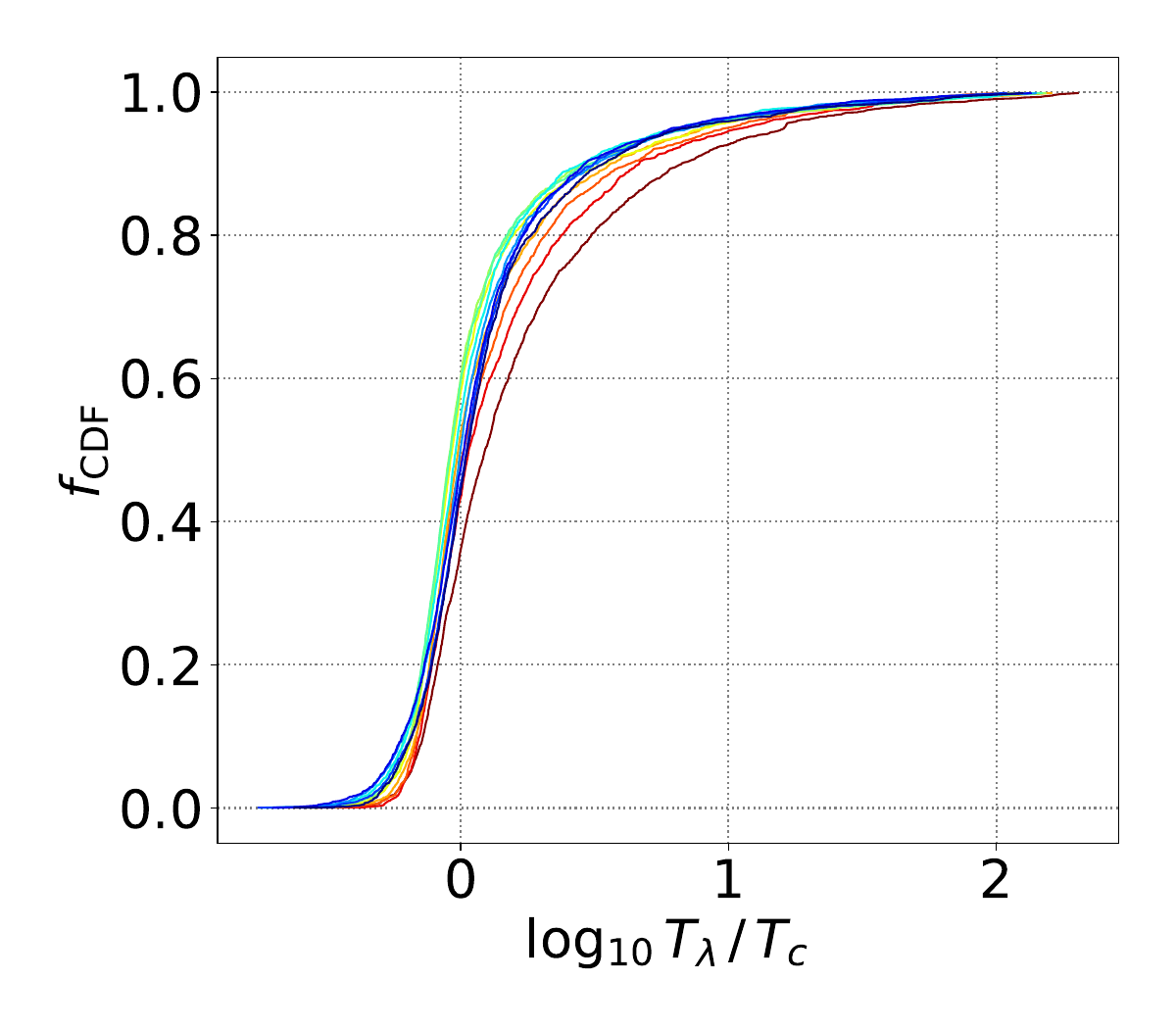} \\
\end{tabular}  
\caption{ Cumulative distribution functions of the amplification factor (left), lifetime (middle) and Lyapunov time scale (right). Triples with larger $L$ have statistically larger amplification factors and lifetimes. For the smallest $L$ values, we observe that $L=0.02$ statistically produces the smallest amplifications rather than $L=0$. Two-sample Kolmogorov-Smirnoff tests between the distributions of $L=0$ and $L=0.19$ give p-values smaller than $10^{-22}$ for each of the three distributions, indicating that all three quantities have a $L$-dependence, including the Lyapunov time scale. Comparing $L=0$ and $L=0.02$ we obtain p-values of 0.00018, 0.0044 and 0.71 (from left to right panel respectively). Hence, the observation that $L=0.02$ produces smaller values of $A$ than $L=0$ is most likely explained by statistically shorter lifetimes, and not by different Lyapunov time scales. We also confirm that the median Lyapunov time scale is of order the crossing time.  }
\label{fig:cdfs}
\end{figure*}

\begin{figure*}
\centering
\begin{tabular}{ccc}

\includegraphics[height=0.24\textwidth,width=0.32\textwidth]{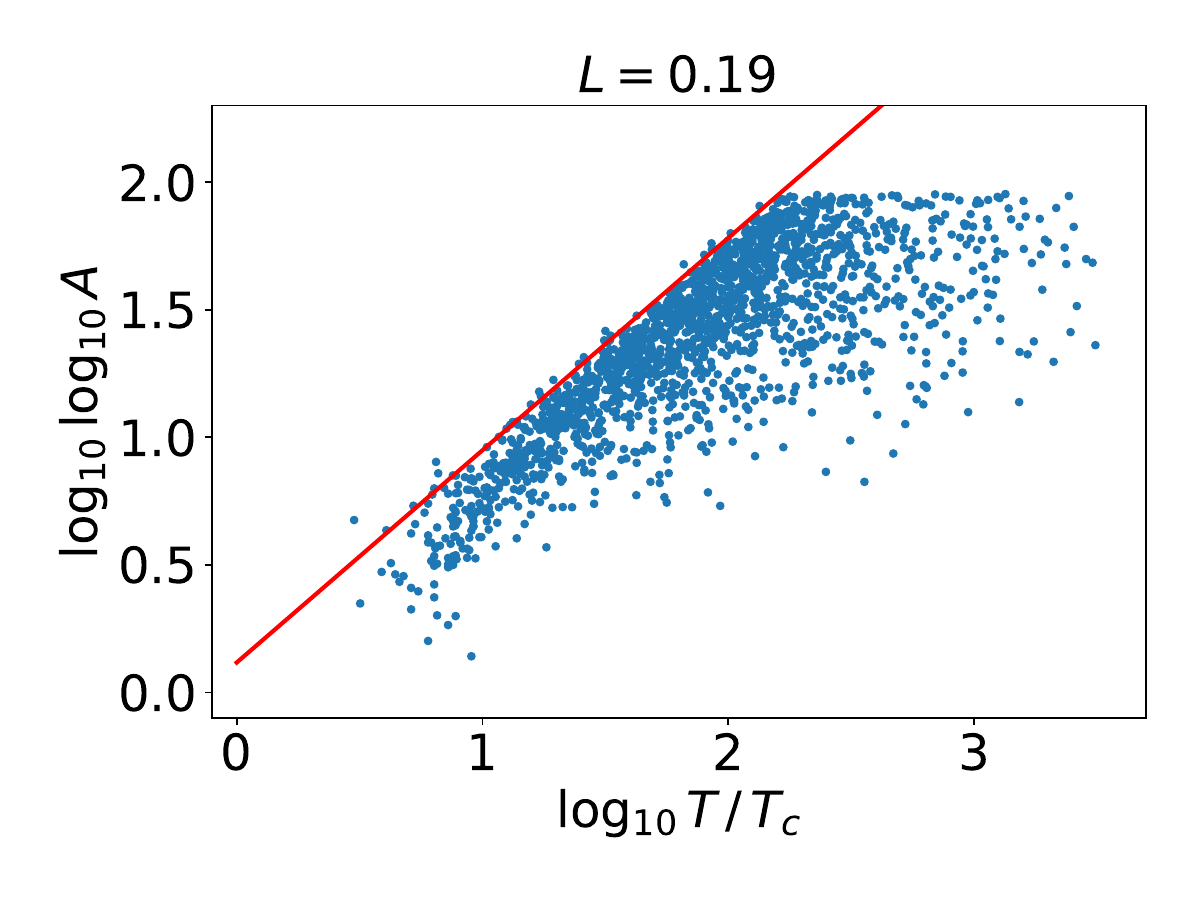} &
\includegraphics[height=0.24\textwidth,width=0.32\textwidth]{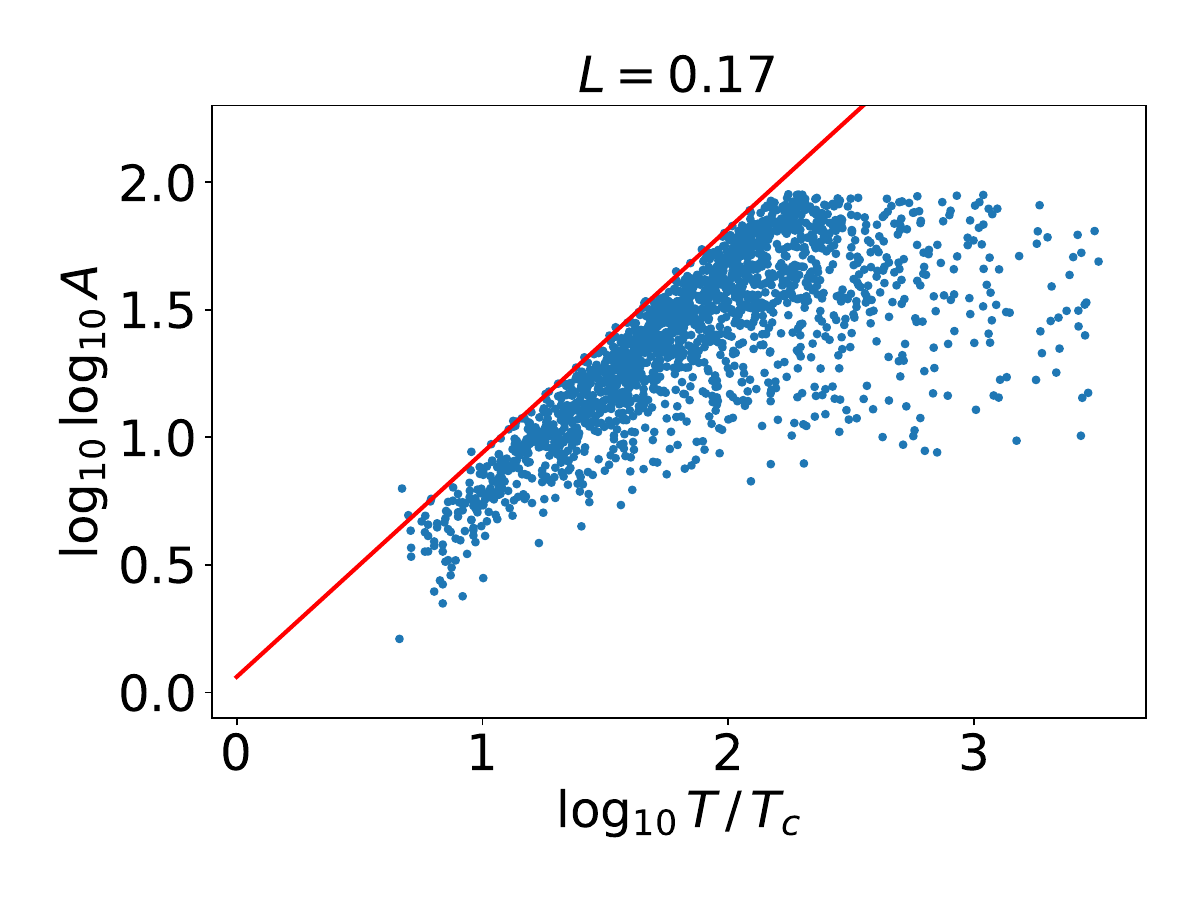} &
\includegraphics[height=0.24\textwidth,width=0.32\textwidth]{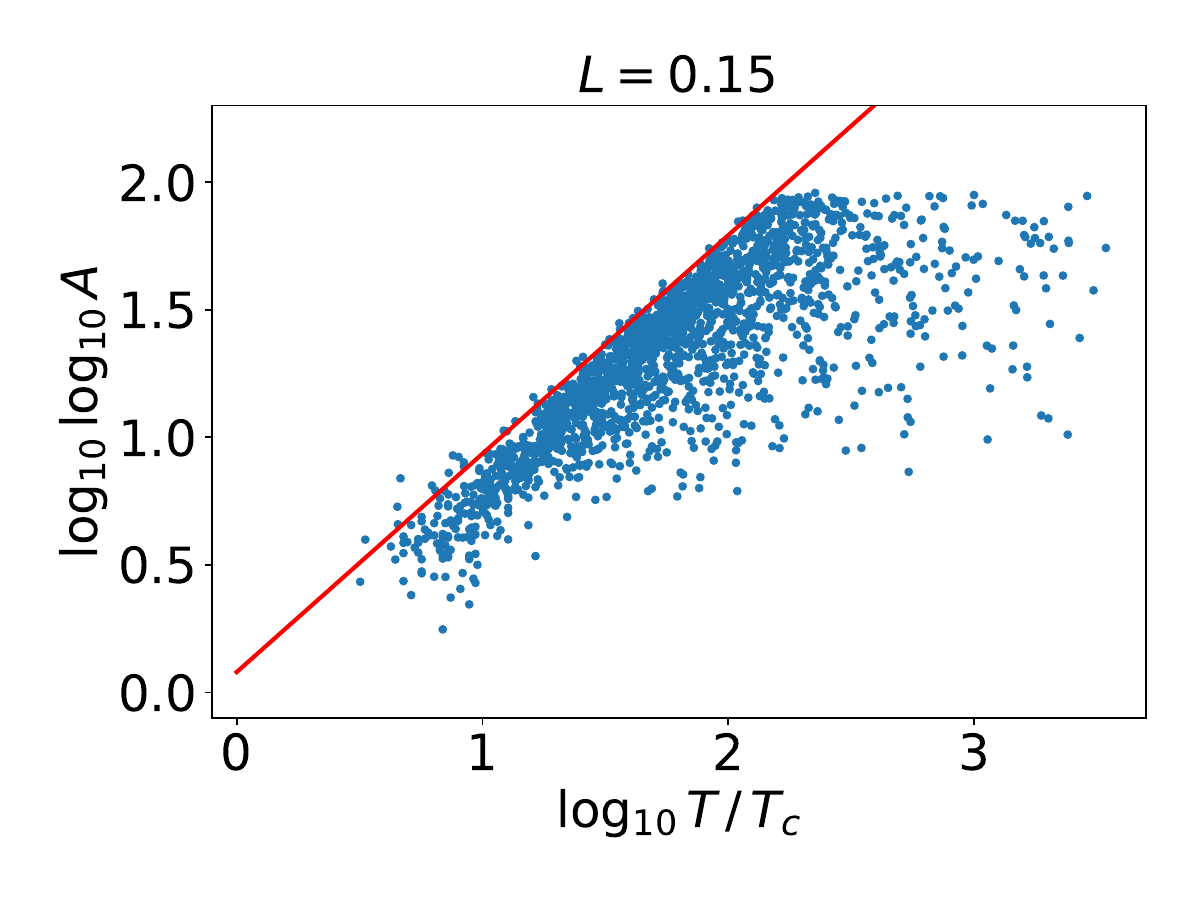} \\

\includegraphics[height=0.24\textwidth,width=0.32\textwidth]{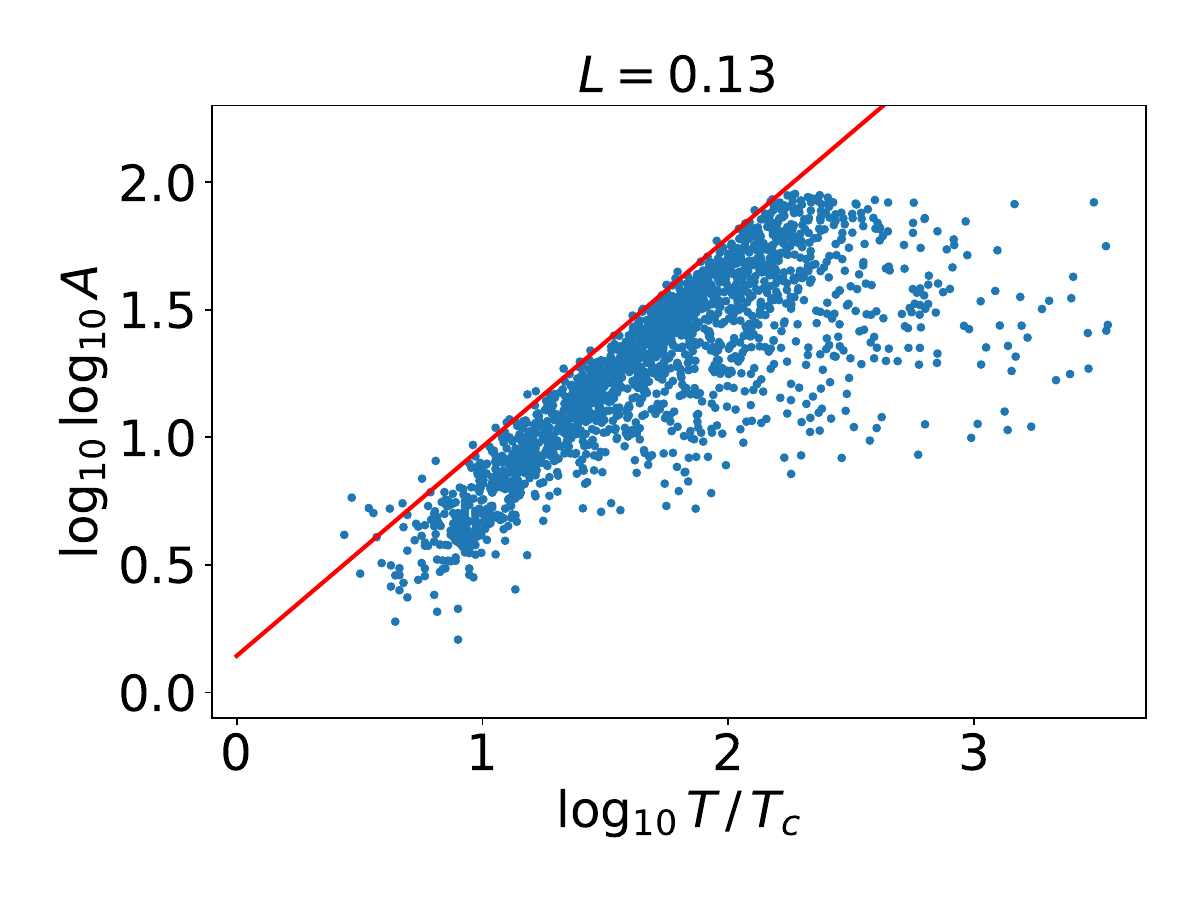} &
\includegraphics[height=0.24\textwidth,width=0.32\textwidth]{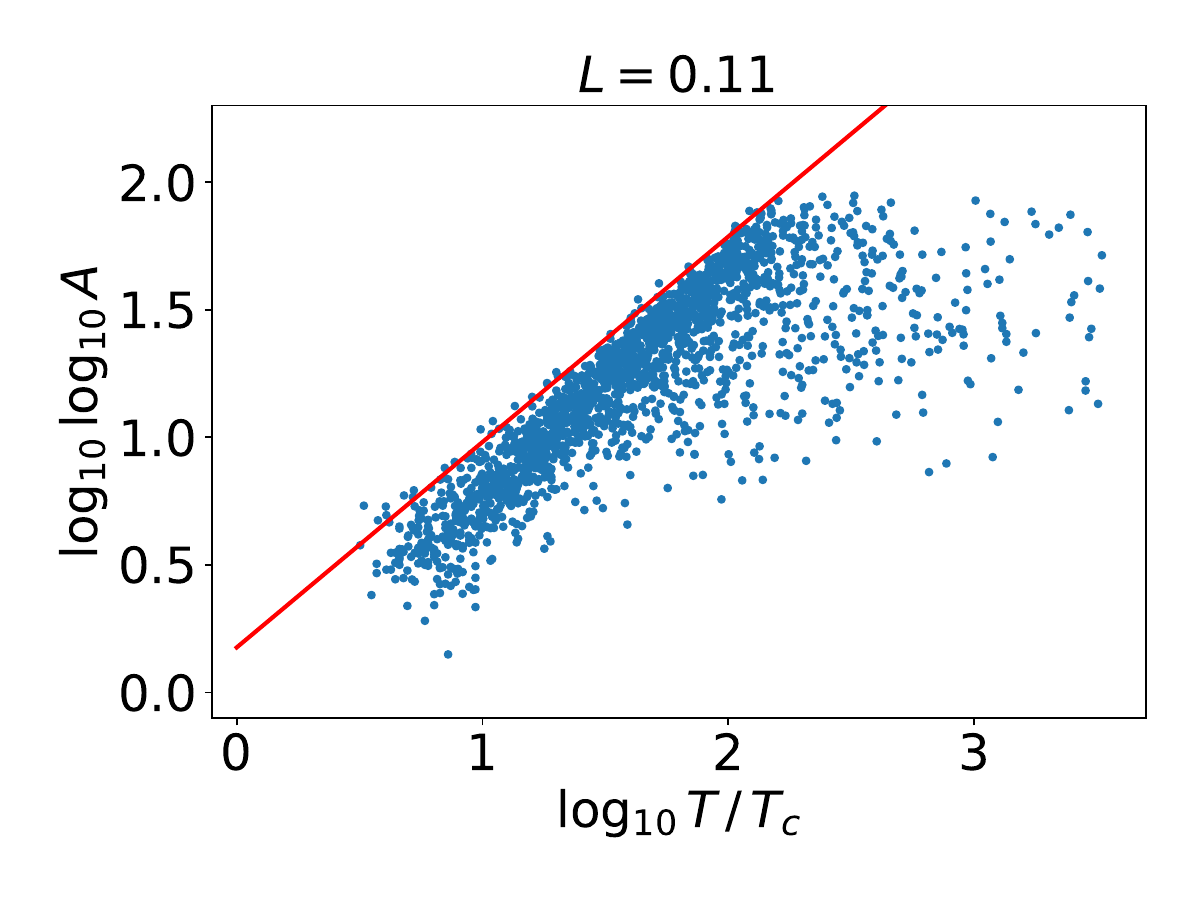} &
\includegraphics[height=0.24\textwidth,width=0.32\textwidth]{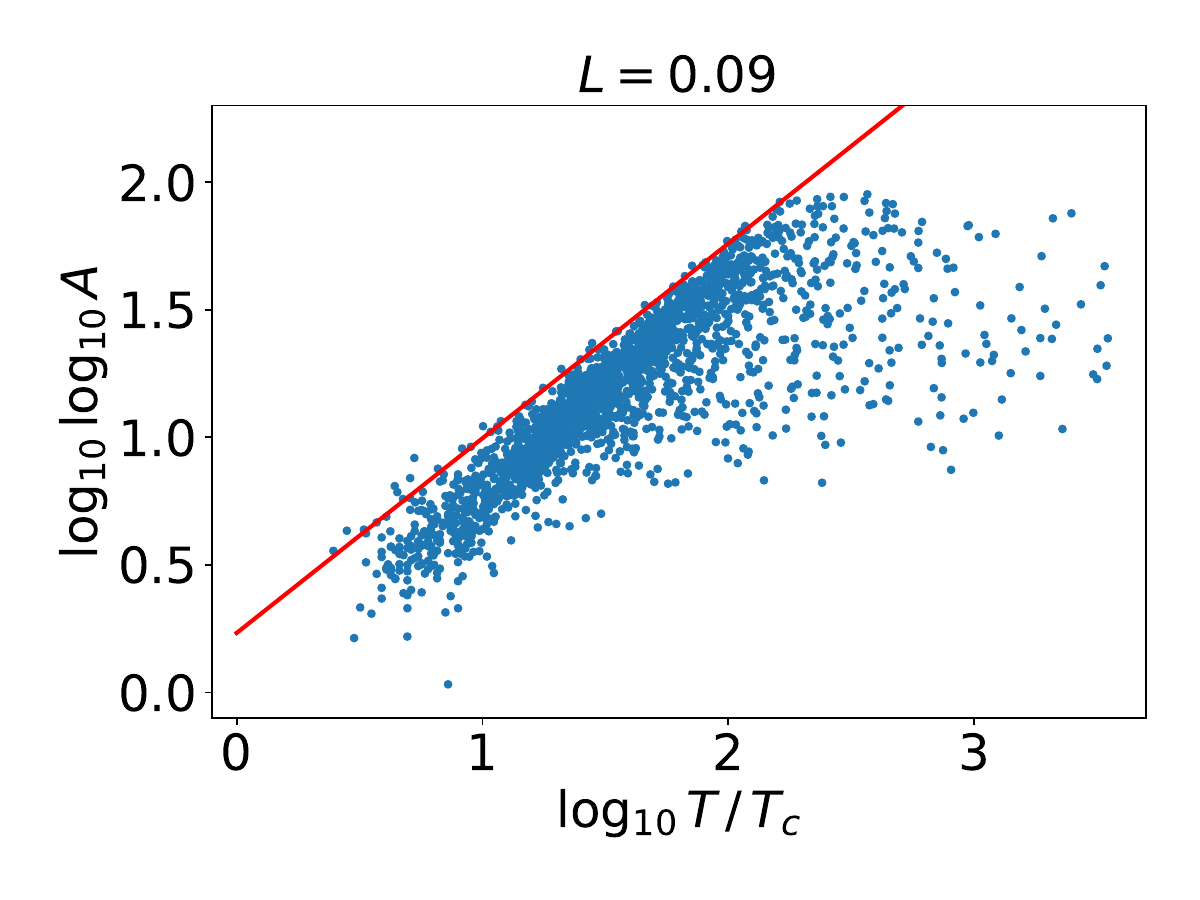} \\

\includegraphics[height=0.24\textwidth,width=0.32\textwidth]{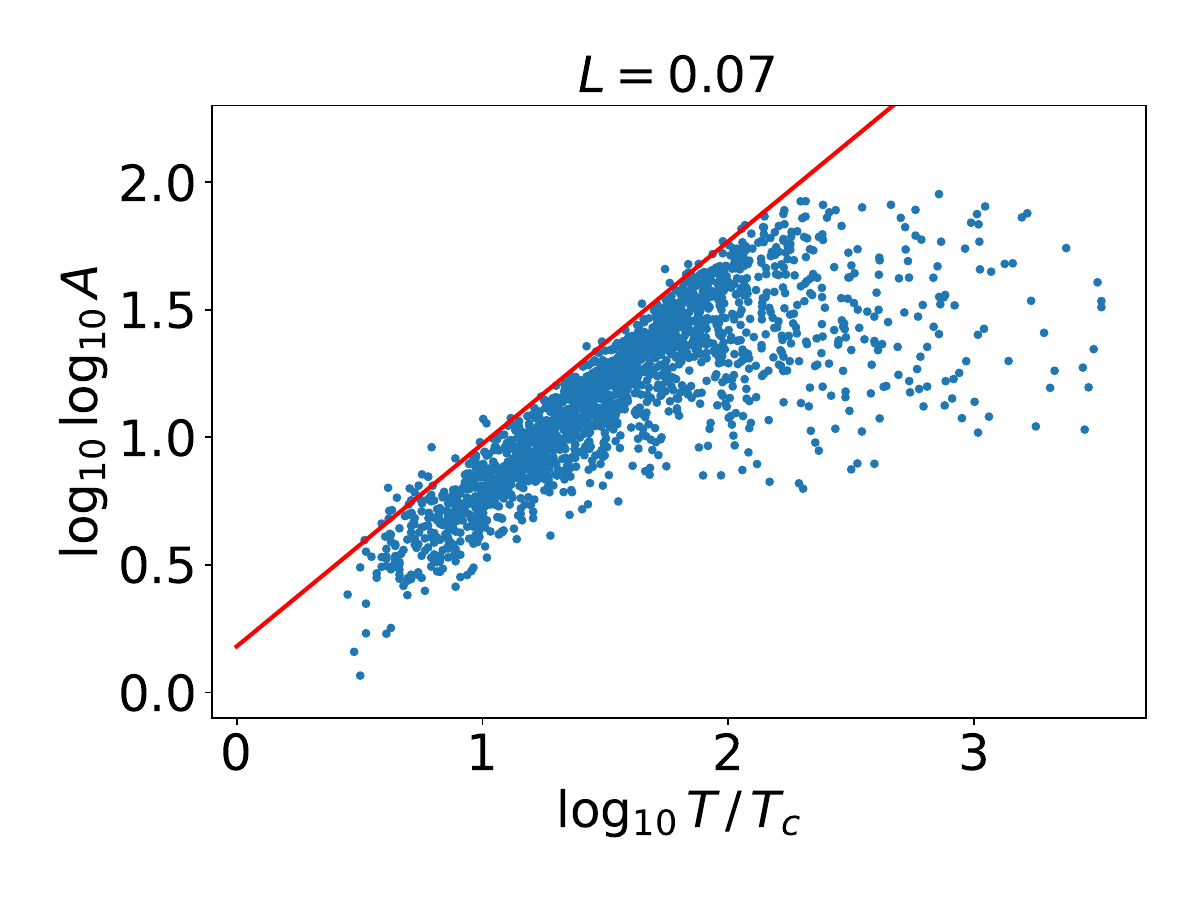} &
\includegraphics[height=0.24\textwidth,width=0.32\textwidth]{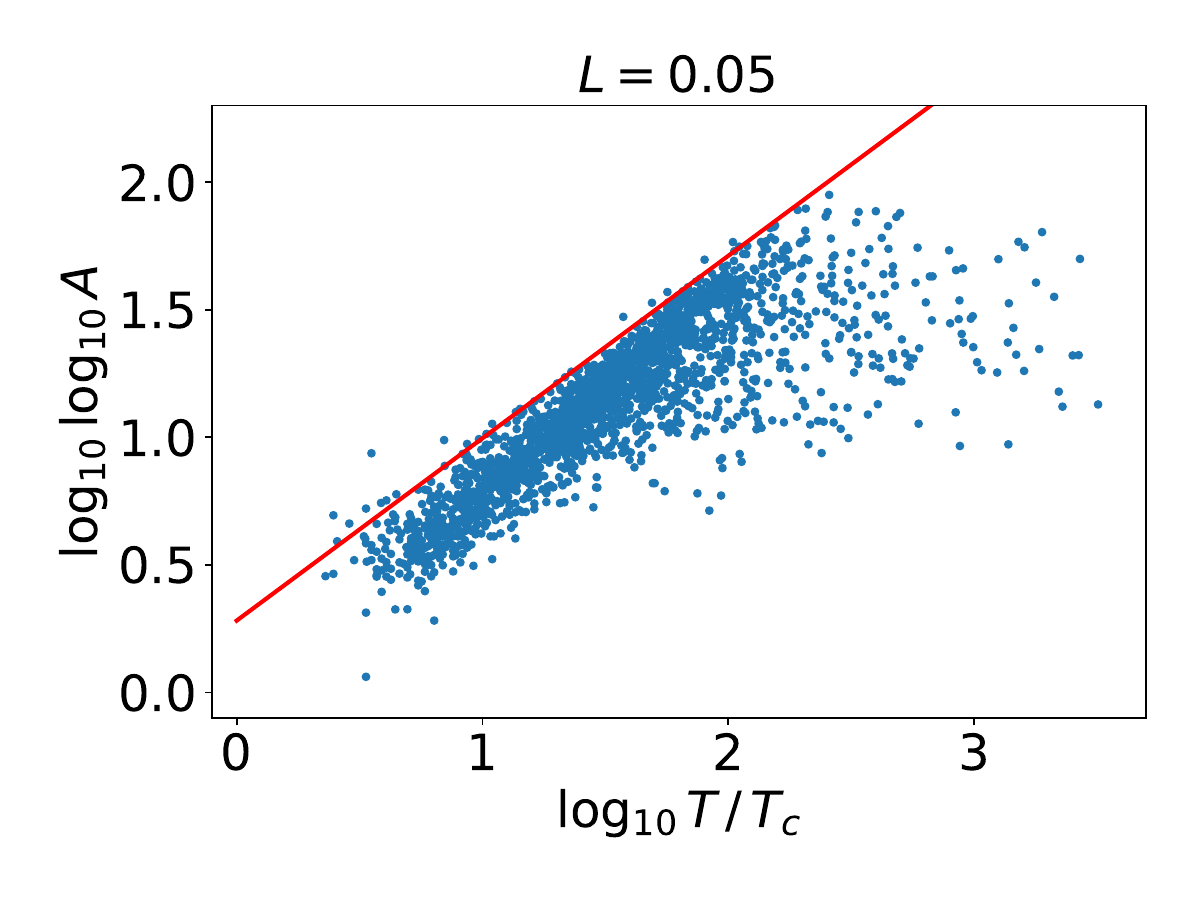} &
\includegraphics[height=0.24\textwidth,width=0.32\textwidth]{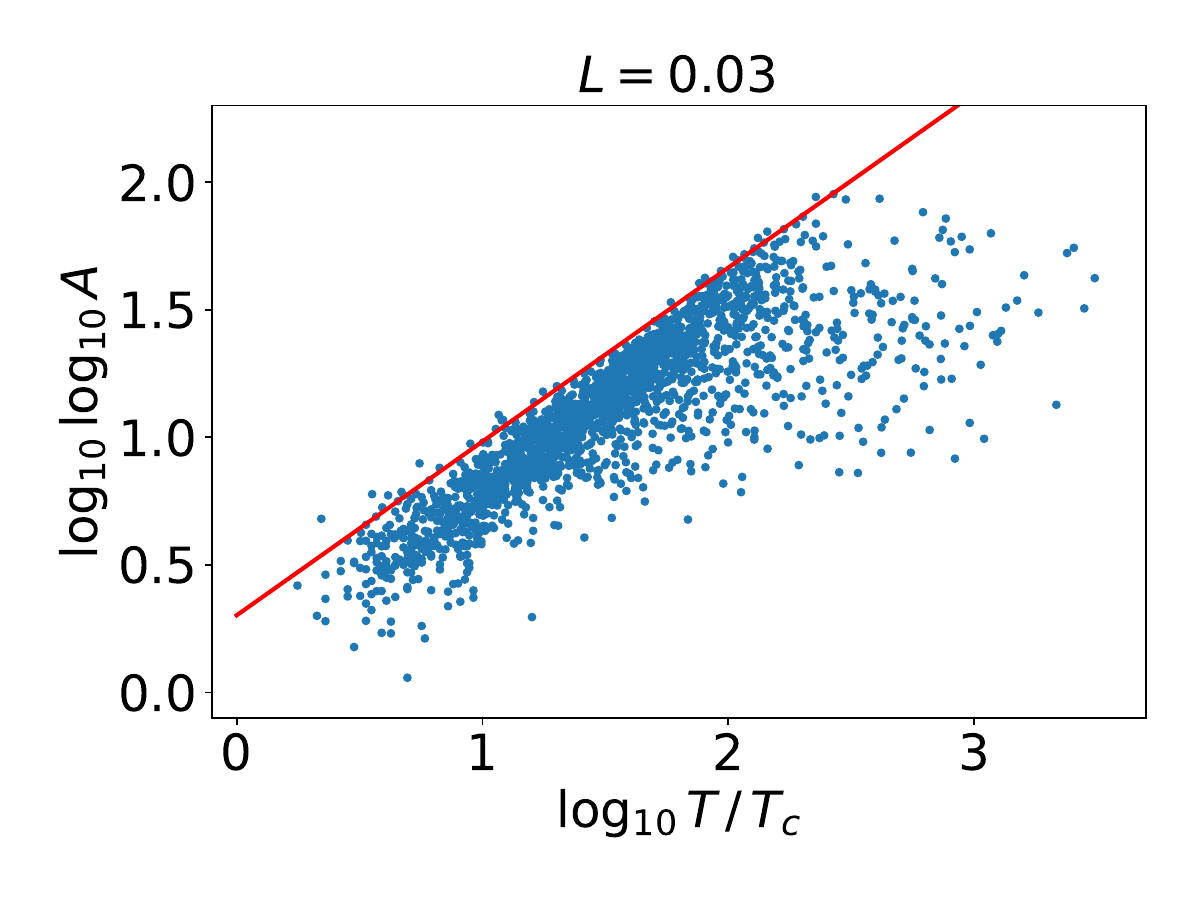} \\

\includegraphics[height=0.24\textwidth,width=0.32\textwidth]{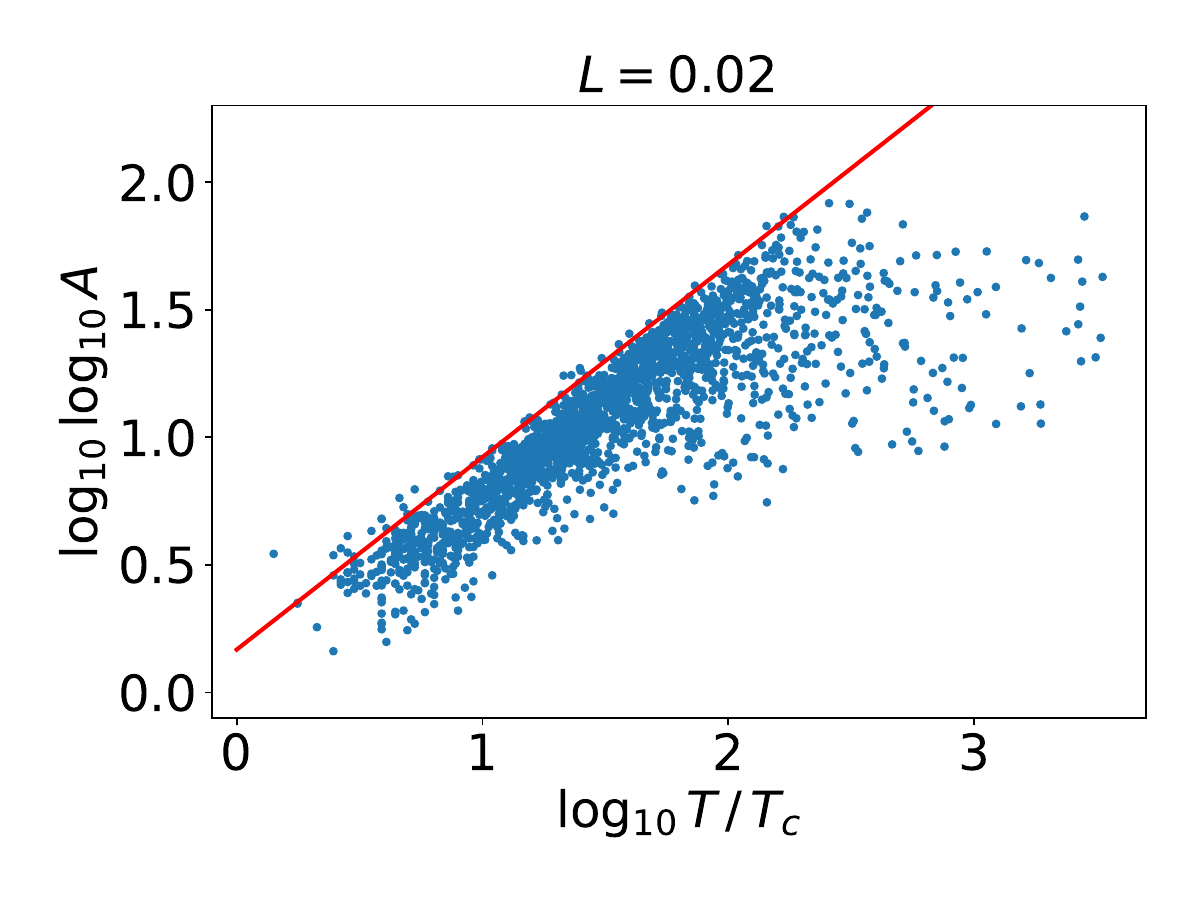} &
\includegraphics[height=0.24\textwidth,width=0.32\textwidth]{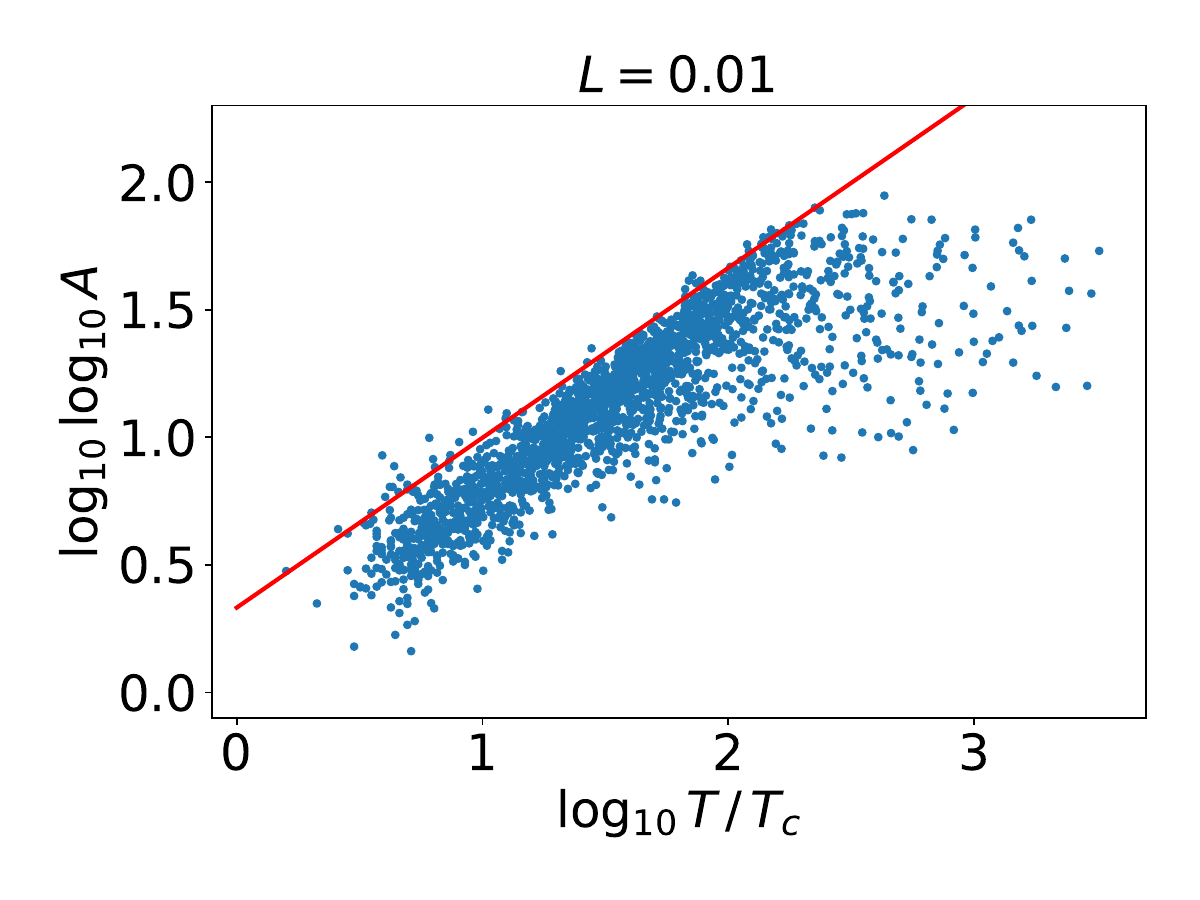} &
\includegraphics[height=0.24\textwidth,width=0.32\textwidth]{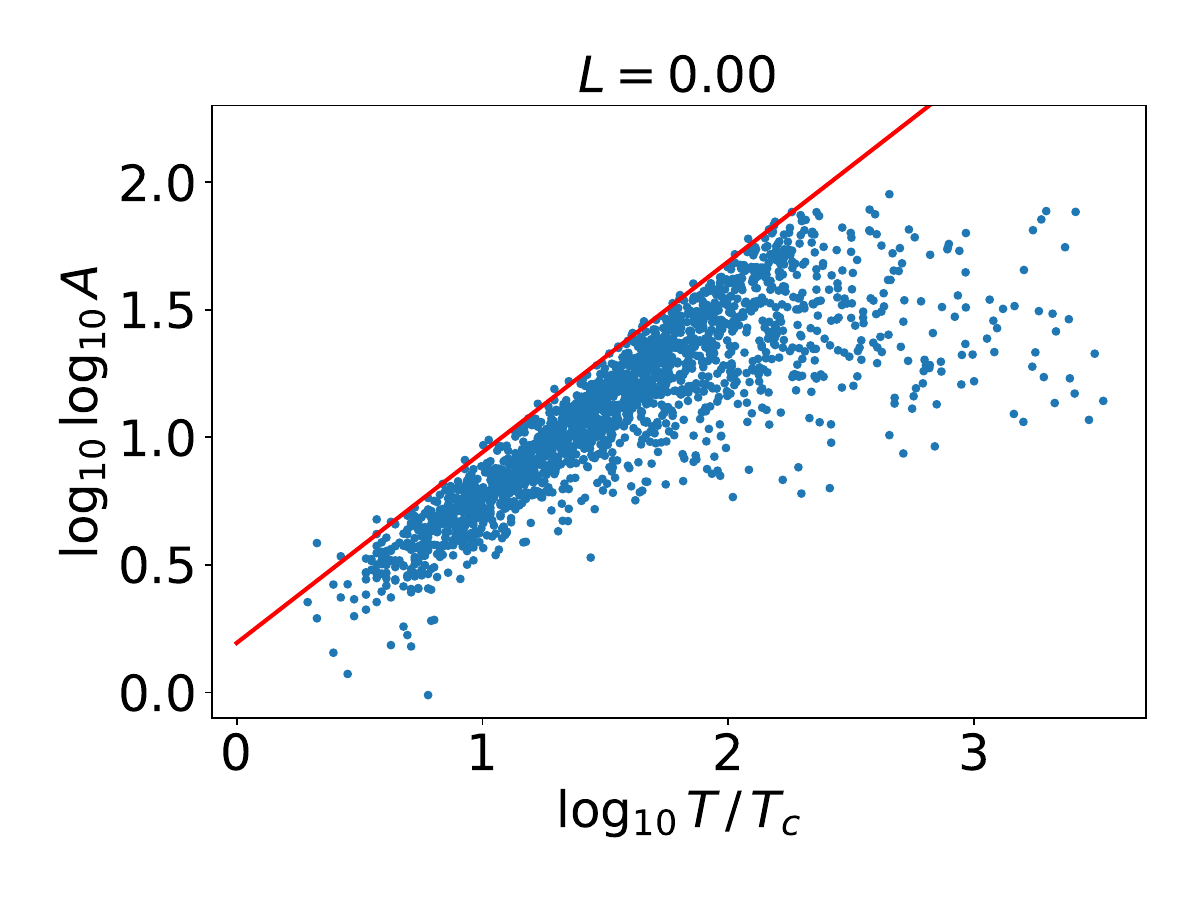} \\

\end{tabular}  
\caption{ Scatter plots of the two main observables: amplification factor, $A$, and lifetime, $T$ (normalised by crossing time, $T_c$). We estimate the slope of the upper edge of the data using linear fits (red lines). The fit parameters are given in Tab.~\ref{tab:ics} with $\alpha$ the slope and $\beta$ the offset. The angular momentum dependence of the slope is visualised in Fig.~\ref{fig:logA_vs_T_slopes}.   }
\label{fig:logA_vs_T_gallery}
\end{figure*}

We first present the outcome of the reversibility experiment in Sec.~\ref{sec:results_rev}. There, we find that ensembles of triples with a higher angular momentum produce a larger fraction of irreversible solutions (see Figs.~\ref{fig:firr} and \ref{fig:firr_diff}). Similar to Paper 1 we find that this is correlated with having larger amplification factors (see Fig.~\ref{fig:cdfs}). In Sec.~\ref{sec:results_logA}, we explore the correlation between the two main observables: amplification factors and lifetimes (see Figs.~\ref{fig:logA_vs_T_gallery}-\ref{fig:logA_vs_T_fits}). 
We will show that higher angular momentum triples can achieve larger amplification factors both due to longer lifetimes as well as larger Lyapunov exponents (see Figs.~\ref{fig:logA_vs_logTc_scatter} and \ref{fig:delta_vs_t_comp}). In Sec.~\ref{sec:discussion} we discuss how the results scale with astrophysical size, and we will also apply our results to tidal perturbations from other bodies in the Universe, which complement intrinsic quantum uncertainties.

\subsection{Reversibility test}\label{sec:results_rev}

In Fig.~\ref{fig:firr}, we plot the fraction of irreversible integrations, $f_{\rm{irr}}$, as a function of numerical accuracy, i.e. Bulirsch-Stoer tolerance, $\epsilon$. For each value of the total angular momentum, we find that the data follows a power law. At low accuracy (large $\epsilon$), $f_{\rm{irr}}$ is of order unity, and with increasing accuracy (smaller $\epsilon$), this fraction decreases according to a power law. The power law index however, depends on angular momentum. We perform linear fits to the data

\begin{equation}
    \log_{10}f_{\rm{irr}} = \nu\log_{10}\epsilon + \eta,
\end{equation}

\noindent where the fit parameters, $\nu$ and $\eta$, can be found in Tab.~\ref{tab:ics}. For the case of $L=0$, we measure a power law index of $\nu = 0.0282 \pm 0.0006$, which is consistent with the measurement of $0.029 \pm 0.001$ for the Agekyan-Anosova map of initial conditions from Paper 1. As we increase the total angular momentum, we find that the power law index first increases to a maximum value of $\nu = 0.033 \pm 0.001$ for $L = 0.02$, and then decreases monotonically until a value of $\nu = 0.0115 \pm 0.0003$ for the initially virialised ensemble. Hence, the easiest triples to reverse have $L = 0.02$, while the hardest triples to reverse are the virial ones with $ L = 0.19$. In Fig.~\ref{fig:firr_diff} we plot the ratio between the largest and smallest fractions from Fig.~\ref{fig:firr}, which are those for $L=0.02$ abd $L=0.19$. It is striking that for $\epsilon = 10^{-90}$, the range in $f_{\rm{irr}}$ among the ensembles is two orders of magnitude, with approximately 10 percent of triples being irreversible still for $L = 0.19$.

In Paper 1 we demonstrated that the fraction of irreversible solutions is determined by the distribution of amplification factors. In Fig.~\ref{fig:cdfs} (left panel), we plot the cumulative distribution function of $\log_{10}A$ for the various ensembles of triples. We confirm that the distributions are angular momentum dependent, such that higher angular momentum triples produce larger amplification factors. We observe that for $L = 0.19$, about 10 percent of triples amplify the initial perturbation by more than a factor of $10^{90}$. Here, we also observe that the ensemble of triples with $L=0.02$ produces the smallest amplifications in a statistical sense. Our observations of the $L$-dependence of $f_{\rm{irr}}$ in Fig.~\ref{fig:firr} are thus reflected in the distributions of $A$ in Fig.~\ref{fig:cdfs} (left panel). 

We also plot the cumulative distribution functions of the lifetime and Lyapunov time in Fig.~\ref{fig:cdfs}. We adopt the two-sample Kolmogorov-Smirnoff (KS) test in order to test whether the various empirical distributions could have been drawn from the same underlying distribution, or whether they are significantly different. We first compare $L=0$ and $L=0.19$ in each of the three panels of Fig.~\ref{fig:cdfs}. The KS test produces p-values smaller than $10^{-22}$ for each of the three distributions, confirming that all three quantities have a statistically significant $L$-dependence. This includes the Lyapunov time scale. 
Whereas the statistical longer lifetimes for $L=0.19$ (middle panel of Fig.~\ref{fig:cdfs}) would contribute towards producing larger amplification factors (left panel), the statistical distribution of Lyapunov times is also shifted towards (somewhat) larger values (right panel), indicating slower growth rates. Nevertheless, the net result is an increase of the amplification factor.  When multiplying the lifetimes by a constant Lyapunov exponent, we find that this does not reproduce the distribution of amplification factors. The $L$-dependence of the amplification factor is thus determined by the $L$-dependence of both the lifetime and the Lyapunov time. We will continue this analysis in the next subsection. 

First, we seek a better understanding of why $L=0.02$ rather than $L=0$ seems to be the easiest ensemble to reverse. We perform KS tests comparing $L=0$ and $L=0.02$ obtaining p-values of 0.00018, 0.0044 and 0.71 for $A$, $T$ and $T_{\rm{\lambda}}$, respectively. At a confidence level of 95\%, we conclude that the distributions of $A$ and $T$ are not drawn from the same underlying distribution, but that $T_{\rm{\lambda}}$ is. Hence, this suggests that $L=0.02$ triples are easier to reverse because they have systematically shorter lifetimes. We leave a physical explanation for this observation in terms of detailed orbital dynamics for follow up studies, although we present some first attempts in Sec.~\ref{sec:results_comparison}. Finally, we observe that the median of the Lyapunov time scale is of order the crossing time (right panel of Fig.~\ref{fig:cdfs}), which is consistent with the theory of punctuated chaos \citep{SPZ23}. 

\subsection{Correlating amplification factors and lifetimes}\label{sec:results_logA}

\begin{figure}
\centering
\begin{tabular}{c}

\includegraphics[height=0.36\textwidth,width=0.48\textwidth]{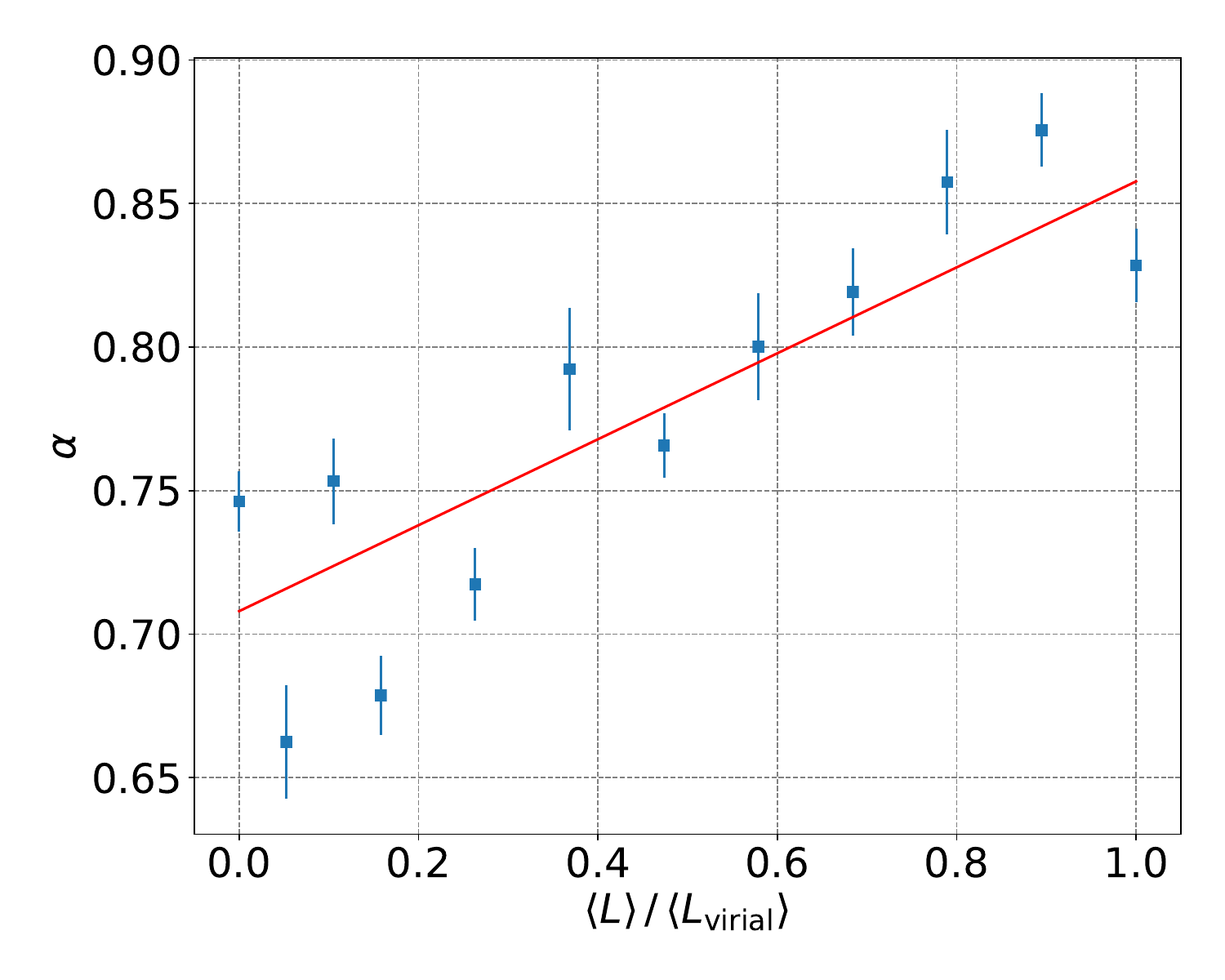} \\

\end{tabular}  
\caption{ Angular momentum dependence of the slope of the linear fits in Fig.~\ref{fig:logA_vs_T_gallery}. A linear fit to the data points (red line) indicates an upward trend. This implies that triples with more angular momentum are able to reach larger amplification factors. This is further visualised in Fig.~\ref{fig:logA_vs_T_fits}. }
\label{fig:logA_vs_T_slopes}
\end{figure}

Our next step is to better understand why high angular momentum triples produce larger amplification factors. 
One contributing factor is that on average high angular momentum triples have longer lifetimes than low angular momentum triples \citep[e.g.][]{2010MNRAS.408.1623O,2015ComAC...2....2B}.
However, we find that simply multiplying the lifetimes of the triples by a constant Lyapunov exponent does not reproduce the measured distribution of amplification factors. Rather than writing

\begin{equation}
    \log A\left( L \right) = \lambda T\left( L \right),
\end{equation}

\noindent with $\lambda$ a constant Lyapunov exponent, we require the more general case given by

\begin{equation}
    \log A\left( L \right) = \lambda\left( T, L \right) T\left( L \right),
    \label{eq:lambda_gen}
\end{equation}

\noindent where the finite-time Lyapunov exponent itself is a function of lifetime and angular momentum \citep[e.g.][]{2007MNRAS.379L..21M, 2009MNRAS.392.1051U}. 

The discussion above motivates the inspection of the correlation between the two main observables in our experiments, namely the amplification factor, $A$, and the lifetime, $T$. In Fig.~\ref{fig:logA_vs_T_gallery}, we present scatter plots of these two quantities for each of our angular momentum ensembles. After experimenting with various combinations of linear and logarithmic axes, we find that a clean linear relation is obtained in the space of $\log \log A$ vs. $\log T$. The first observation we make is that for each given $T$, there is a (sharp-edged) maximum value of $A$. This diagonal upper ridge continues up to about $100\,T_c$, after which $A$ flattens. By this time, $A$ has reached a value of $10^{90}$, which is the limit in our experiment. If we had continued to decrease the Bulirsch-Stoer tolerance beyond $10^{-90}$, we expect the diagonal trend to continue. Hence, if we only consider the resolved portion of the scatter plot ($T \leq 100\,T_c$), we observe a sharp upper ridge in the data, which is indicative of a maximum Lyapunov exponent, i.e. a maximum rate of divergence. 
The second observation is the gradual scatter of data points towards very long lifetimes. The maximum lifetime in our experiment was set to $10^4$ H\'enon time units ($\sim3536\,T_c$). There is a large range of lifetimes, which still end up with the same amplification factor. This observation can be interpreted by stating that the evolution of long-lived triples is driven by prolonged excursions of a single body during which it is only weakly interacting with the binary system. The exponential sensitivity during these phases is greatly reduced (see also Fig.~\ref{fig:delta_vs_t_comp}). 

\begin{figure}
\centering
\begin{tabular}{c}

\includegraphics[height=0.36\textwidth,width=0.48\textwidth]{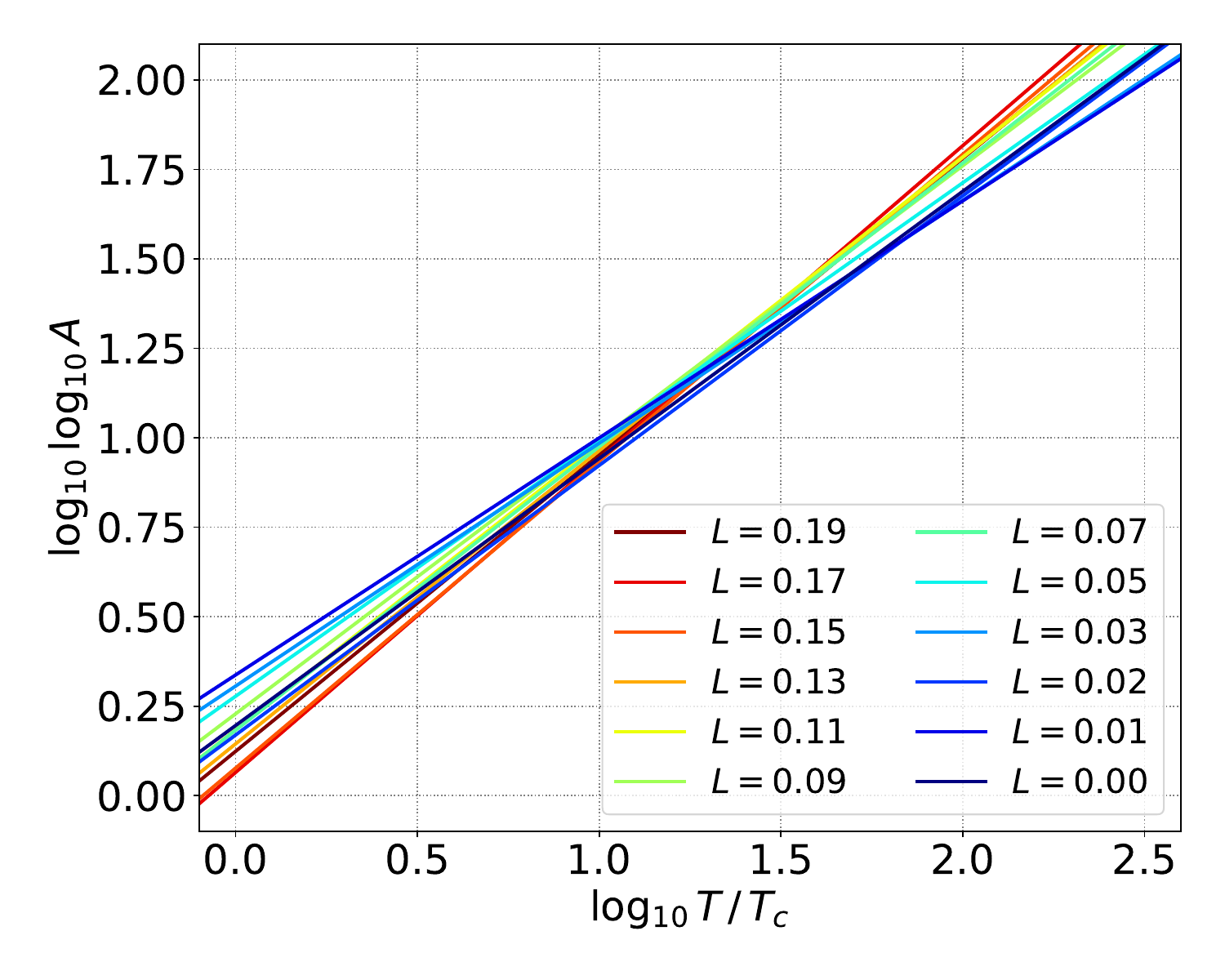} \\

\end{tabular}  
\caption{ We collect the linear fits from Fig.~\ref{fig:logA_vs_T_gallery} for a direct comparison. Initially, up to about 20 crossing times, larger amplification factors are reached by the low angular momentum triples. At longer lifetimes however, the high angular momentum triples take over.    
}
\label{fig:logA_vs_T_fits}
\end{figure}

Coming back to the first observation, we fit the upper edge of the resolved data with a linear model,

\begin{equation}
    \log_{10}\log_{10}A = \alpha \log_{10}\frac{T}{T_c} + \beta,
    \label{eq:fit}
\end{equation}

\noindent where the fitting parameters, $\alpha$ and $\beta$, are given in Tab.~\ref{tab:ics}. Here, we made bins along the horiztonal axis, and used a bootstrap resampling method to estimate error margins to the fitting parameters. 
In Fig.~\ref{fig:logA_vs_T_slopes}, we plot the slopes of the linear fits, $\alpha$, as a function of angular momentum: high angular momentum triples have a steeper slope, i.e. $\alpha$ increases from $\sim0.70$ to $\sim0.86$. The implication is best observed in Fig.~\ref{fig:logA_vs_T_fits}. There, we find that up to about $20\,T_c$ the largest amplification factors are reached by the lower angular momentum triples. At longer times however, the higher angular momentum triples become dominant, e.g. larger values of $A$ by $\sim 14$ orders of magnitude for $T\,/\,T_c = 100$. 
This result is more in line with the common intuition that dynamically cold triples have a violent and short life, while virial triples are relatively calm and gradually build up their chaos in the long term.

\begin{figure*}
\centering
\begin{tabular}{c}

\includegraphics[height=0.8\textwidth,width=0.96\textwidth]{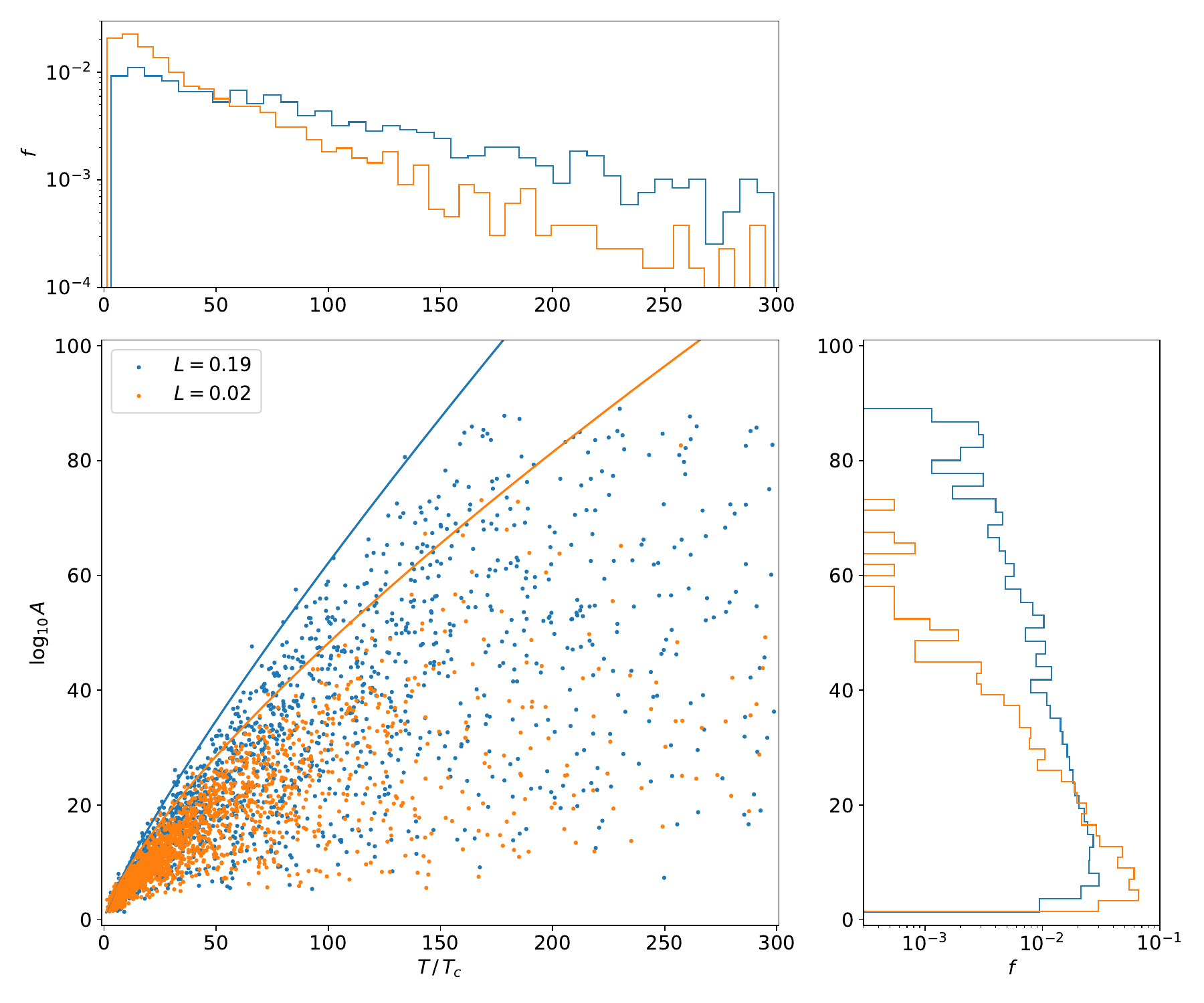} \\

\end{tabular}  
\caption{ Relating the two main observables of the reversibility test: amplification factor, $A$, and lifetime, $T$. We compare the two most different ensembles given by $L=0.02$ (orange) and $L=0.19$ (blue). We observe a bias both in the lifetime distribution (top histogram), and in the maximum achievable value of $A$ given a certain value of $T$ (scatter plot). The combination of longer lifetimes and larger growth rates leads to a larger fraction of high amplification factors (right histogram), and thus also a higher fraction of irreversible solutions (see Fig.~\ref{fig:firr}).   }
\label{fig:logA_vs_logTc_scatter}
\end{figure*}

As a further illustration of the influence of angular momentum, we provide a detailed comparison between low ($L = 0.02$) and high ($L = 0.19$) angular momentum triples in  Fig.~\ref{fig:logA_vs_logTc_scatter}. In the lifetime histogram (top panel), we observe that a higher angular momentum indeed results in a larger fraction of long-lived triples. In the main scatter plot, we also plot the corresponding fits to the upper ridge line in the data (from Figs.~\ref{fig:logA_vs_T_gallery}-\ref{fig:logA_vs_T_fits}). Since the slope in this plot corresponds directly to the Lyapunov exponent ($\lambda \propto \log A\,/\,T$), we notice that $\lambda$ has a dependence on $L$, and since the slope in each curve flattens towards larger lifetimes, $\lambda$ also depends on $T$. We have thus confirmed the general case in Eq.~\eqref{eq:lambda_gen}, where $\lambda = \lambda\left( L, T \right)$. 
The combination of longer lifetimes and larger Lyapunov exponents causes higher angular momentum triples to achieve increasingly larger amplification factors (see histogram in the right panel of Fig.~\ref{fig:logA_vs_logTc_scatter}).

We finish this subsection by deriving a relation between lifetime and Lyapunov time. Using Eq.~\ref{eq:fit}, we can write

\begin{equation}
    \log A = 10^\beta \left( \frac{T}{T_c} \right)^\alpha.
\end{equation}

\noindent Dividing both sides by $T\,/\,T_c$ we obtain

\begin{equation}
    \lambda T_c = 10^\beta \left( \frac{T}{T_c} \right)^{\alpha-1},
\end{equation}

\noindent or in terms of Lyapunov time:

\begin{equation}
    \frac{T_\lambda}{T_c} = 10^{-\beta} \left( \frac{T}{T_c} \right)^{1-\alpha}.
\end{equation}

\noindent Rewriting for $T$ gives:

\begin{equation}
    \frac{T}{T_c} = 10^{\frac{\beta}{1-\alpha}} \left( \frac{T_\lambda}{T_c} \right)^{\frac{1}{1-\alpha}}.
    \label{eq:t_vs_tl}
\end{equation}

\noindent This expression describes a power law relation between lifetime and Lyapunov time, corresponding to the maximum growth rate. The $L$-dependent power law index of $T_\lambda$ ranges from 3 to 8, which is much steeper than estimates from previous studies \citep[e.g.][]{2007MNRAS.379L..21M, 2009MNRAS.392.1051U}, although they focused on the median growth rate rather than the maximum rate. 
From Eq.~\ref{eq:t_vs_tl} it may seem that triples with a shorter Lyapunov time scale have shorter lifetimes. However, this statement assumes $T_\lambda$ is a constant, such that if it were possible to measure a triple's instantaneous Lyapunov time scale, one could predict its lifetime. However, this becomes problematic due to the large time variations in the rate of divergence, as we will discuss next. 
 
\subsection{Rate of divergence as a function of time. }\label{sec:results_comparison}

\begin{figure*}
\centering
\begin{tabular}{c}
\includegraphics[height=0.66\textwidth,width=0.88\textwidth]{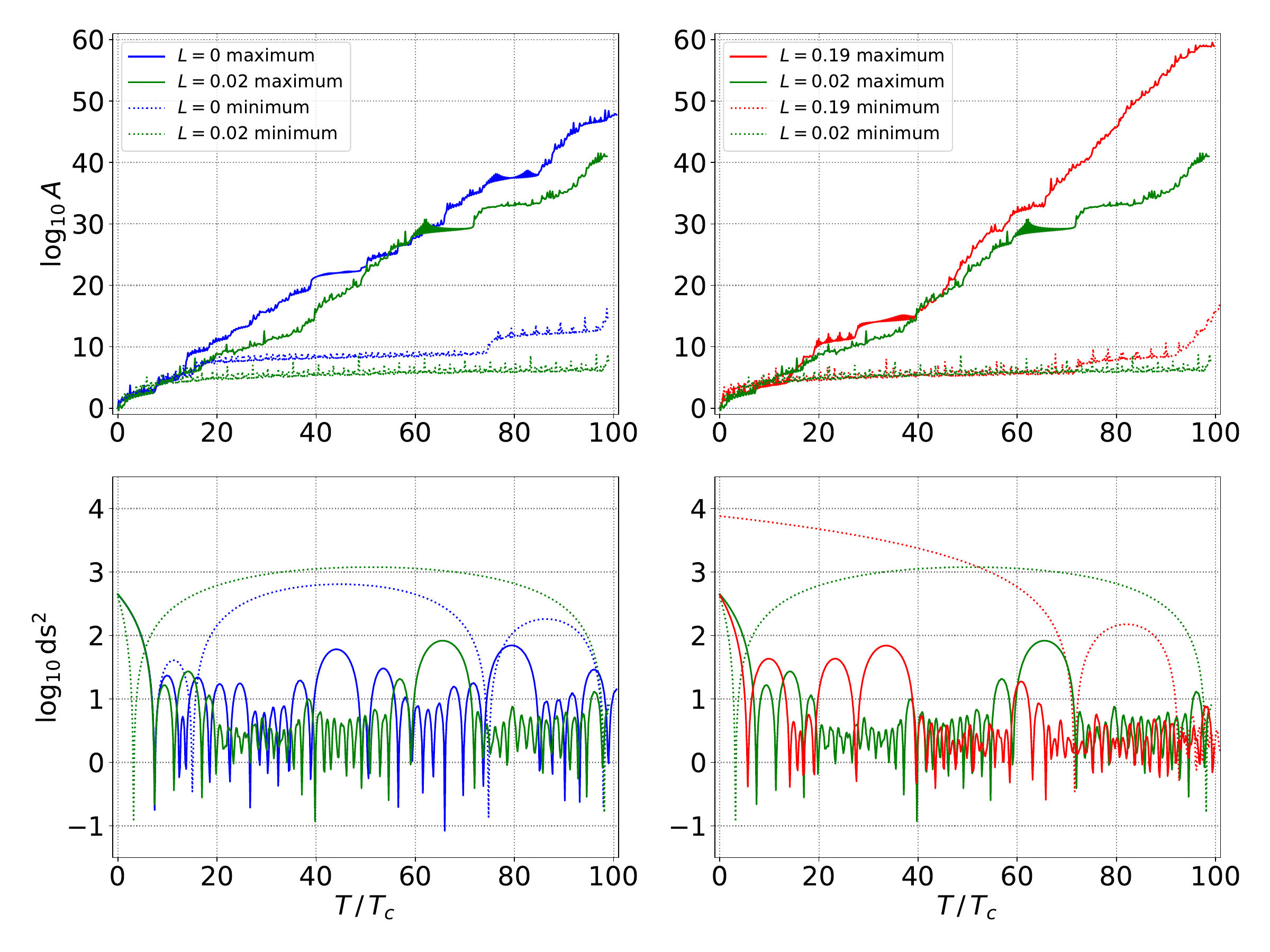} \\
\end{tabular}  
\caption{For triple systems with a lifetime of about 100 crossing times, we gather the solutions with the largest and smallest amplification factor. We plot the time evolution of the amplification factor (top panels), and of the metric $\rm{ds}^2$ (bottom panels), which is the sum of the squared distances between all pairs. We compare the low angular momentum ensembles $L=0$ and $L=0.02$ (left column), and the most and least chaotic ensembles $L=0.19$ and $L=0.02$ (right column). Rapid growth is driven by democratic resonant interactions (high frequency oscillations in $\rm{ds}^2$), while slow growth corresponds to long excursions of a single body from the bound pair.  }
\label{fig:delta_vs_t_comp}
\end{figure*}

In order to reach a deeper understanding of the physical origin of the empirical relations measured so far, it becomes necessary to closely correlate the instantaneous Lyapunov exponents to the orbital configurations. How do the triple components drive the growth of perturbations, and in general, how does that depend on angular momentum, mass ratio etc.? Although these are challenging but important questions requiring follow up studies, here we perform an analysis in the style of \citet{1986LNP...267..212D}. They define the metric

\begin{equation}
    \rm{ds}^2 = \sum_{i=1}^3 \sum_{j>i} \left( x_j - x_i \right)^2 + \left( y_j - y_i \right)^2 + \left( z_j - z_i \right)^2,
\end{equation}

\noindent which is the sum of the squared distances between every pair of bodies. This metric is only small if all three bodies are close together, and during such moments amplifications are expected to be strong. On the other hand, if the metric attains very large values, then this corresponds to an excursion of one of the bodies from the bound pair. During such excursions the amplification is expected to be small. 
From our ensembles of simulations, we gather solutions with a fixed lifetime of $100 \pm 2$ crossing times. Then, for each ensemble of $L$ we select the solutions with the maximum and minimum amplification factor, i.e. the most and least chaotic solutions. We will make a comparison of these solutions between the most and least chaotic ensemble, $L=0.19$ vs. $L=0.02$, and also between the low angular momentum ensembles, $L=0$ vs. $L=0.02$. In Fig.~\ref{fig:delta_vs_t_comp} we compare the various solutions, and correlate the time evolution of the amplification factor with the metric defined above.  

The metric shows an oscillatory behaviour indicating that the size of the triple itself oscillates. For the least chaotic solutions (those with the minimum values of $A$), we indeed observe that the metric attains large values (over a 100) and over an extended period (over 40 crossing times). The least chaotic solutions clearly correspond to very long excursions, during which the growth is linear. When the metric reaches a minimum value however, we observe a corresponding jump in $A$. This behaviour is consistent with punctuated chaos in which the rate of divergence is linear over extended times, but punctuated by big brief jumps due to strong events in the system, such as close encounters. According to this theory a higher frequency of events results in a more rapid amplification. The most chaotic solutions with the maximum values of $A$ indeed correspond to a much higher oscillation frequency of the metric. 

Sustained exponential growth is thus achieved if long excursions are absent, i.e. there is a prolonged democratic resonant interaction among the bodies. Comparing the most rapidly growing solutions for the three different values of $L$, we find that the number of events (i.e. minima of the metric) is not proportional to the rate of growth. For the most chaotic solution with $L=0.19$ we count 50 oscillations, while for the least chaotic case $L=0.02$ we count a similar 48 oscillations. For $L=0$ we only count 28 oscillations,  while still achieving a larger value of $A$ than $L=0.02$. Hence, we find that even during democratic resonant phases the growth is variable, indicating that other metrics also play a role. For example, considering the maximum $L=0.02$ solution, we observe a rapid exponential growth between $40 \leq T/T_c \leq 50$, and a rather slow growth during $80 \leq T/T_c \leq 90$, even though both correspond to phases of high frequency of the metric with the same value of $L$.

The driver of chaos is therefore not (solely) close encounters in radial orbits, but rather the prolonged and non-linear interaction among all three bodies
in a democratic configuration. Based on our numerical results, we
speculate that high angular momentum triples tend to have a longer
cumulative resonant interaction time and/or shorter excursion phases
of a single body, thereby effectively reducing the phases of slow growth. 
However, even during democratic three-body interaction phases the rate of divergence depends on other factors besides the metric $\rm{ds}^2$, including $L$. 
This motivates a closer inspection of the dependence
of the instantaneous Lyapunov exponent on the specifics of the orbital
configuration in follow up studies.

\section{Discussion and Conclusions}\label{sec:discussion}

\subsection{Angular momentum dependence}

We revisit the astrophysical application of Paper 1, which considered three supermassive black holes with a mutual separation of order one parsec. In the zero angular momentum limit, we confirm their result that about 5\% of triples are irreversible up to the Planck length (see Tab.~\ref{tab:ics}). This result is robust with respect to the type of initial condition (Plummer or Agekyan-Anosova map), as the most chaotic triples tend to forget their specific initial condition, and their ultimate fate is determined by global conserved quantities, i.e. angular momentum and energy. 

Depending on the astrophysical context, triples with zero angular momentum might only represent a very small fraction of the population. Furthermore, one might naively expect that triples with a higher angular momentum might be less chaotic. This intuition mainly stems from the idea that chaos is driven by close encounters, which are more likely to occur in low angular momentum systems with radial orbits. Hence, it was expected that the 5\% of fundamentally unpredictable triples would be (greatly) reduced when considering a more general and realistic population of triples with varying, non-zero angular momenta. However, our results demonstrate the contrary; fundamentally unpredictable triples exist over a wide range of angular momenta, and the fraction of fundamentally unpredictable triples is even enhanced up to about 30\% of the population (for the initially virialised case). 

We speculate that hierarchical triple systems near the edge of stability \citep[e.g.][]{Toonen2022}, which can dynamically break up, also include a fraction of fundamentally unpredictable systems, similar to our highest angular momentum case. It is obvious that if the hierarchy of the triple configuration is increased, that the effect of the tertiary is diminished, thereby reducing the degree of exponential sensitivity. The transition from order to chaos is at the basis of some stability criteria for hierarchical triple systems \citep[e.g.][]{1999ASIC..522..385M, Nik2008}. 

\subsection{Dependence on astrophysical scale}\label{sec:scale}

\begin{table*}
\centering
\begin{tabular}{| l | l | l | l | l | l | l } 
\hline
Triple System & Mass scale [kg] & Length scale [m] & Crossing time [yr] & $\log_{10}A_{\rm{min}}$ & $p_{\rm{h}}[\%]$ & $p_{\rm{star}}[\%]$ \\
\hline\hline
Supermassive black holes       & $2.0 \times 10^{36}$ & $3.1 \times 10^{16}$ & $4.2 \times 10^4$ & 51.3 & [0.031, 0.30] & [0.49, 0.77] \\ 
Stars       & $2.0 \times 10^{30}$ & $7.0 \times 10^{12}$ & $1.4 \times 10^2$ & 47.6 & [0.040, 0.33] & [0.33, 0.68] \\ 
Jupiters    & $2.0 \times 10^{27}$ & $1.5 \times 10^{11}$ & $1.4 \times 10^1$ & 46.0 & [0.046, 0.34] & [0.29, 0.64] \\ 
Moons       & $1.0 \times 10^{21}$ & $1.0 \times 10^8$ & $3.5 \times 10^{-1}$ & 42.8 & [0.058, 0.37] & [0.22, 0.59] \\ 
Asteroids   & $1.0 \times 10^{19}$ & $7.7 \times 10^7$ & 2.3 & 42.7 & [0.059, 0.37] & [0.25, 0.62] \\ 
Pebbles     & 1.0 & 3.3 & $6.6 \times 10^{-2}$ & 35.3 & [0.10, 0.45] & [0.20, 0.57] \\ 
Dust grains & $1.0 \times 10^{-4}$ & $1.0 \times 10^{-3}$ & $3.5 \times 10^{-5}$ & 31.8 & [0.14, 0.49] & [0.12, 0.48] \\ 
\hline
\end{tabular}
\caption{ Fraction of fundamentally unpredictable triples for various astrophysical scales. For each system we give the characteristic mass and length scale, as well as the corresponding value of the dynamical crossing time. The minimum amplification factor required for a Planck length fluctuation to reach the size of the triple ($A_{\rm{min}}$) is also given. The fractions of irreversible systems are given by: $p_h$, for the intrinsic Planck length fluctuations (see Fig.~\ref{fig:multi_scale}), and $p_{\rm{star}}$, for the tidal perturbation from a star at a distance of 1\,kpc (see Fig.~\ref{fig:isolation}). Each fraction is given as a range, where the lower limit is obtained from the $L=0.02$ ensemble, and the upper limit from the virial $L=0.19$ ensemble. We find that compact triples are more susceptible to Planck length perturbations, while tidal perturbations become increasingly important for larger triples. On the scale of dust grains, we find that both sources of perturbations become similar in magnitude.   }
\label{tab:triple_scale}
\end{table*}

\begin{figure*}
\centering
\begin{tabular}{c}

\includegraphics[height=0.48\textwidth,width=0.64\textwidth]{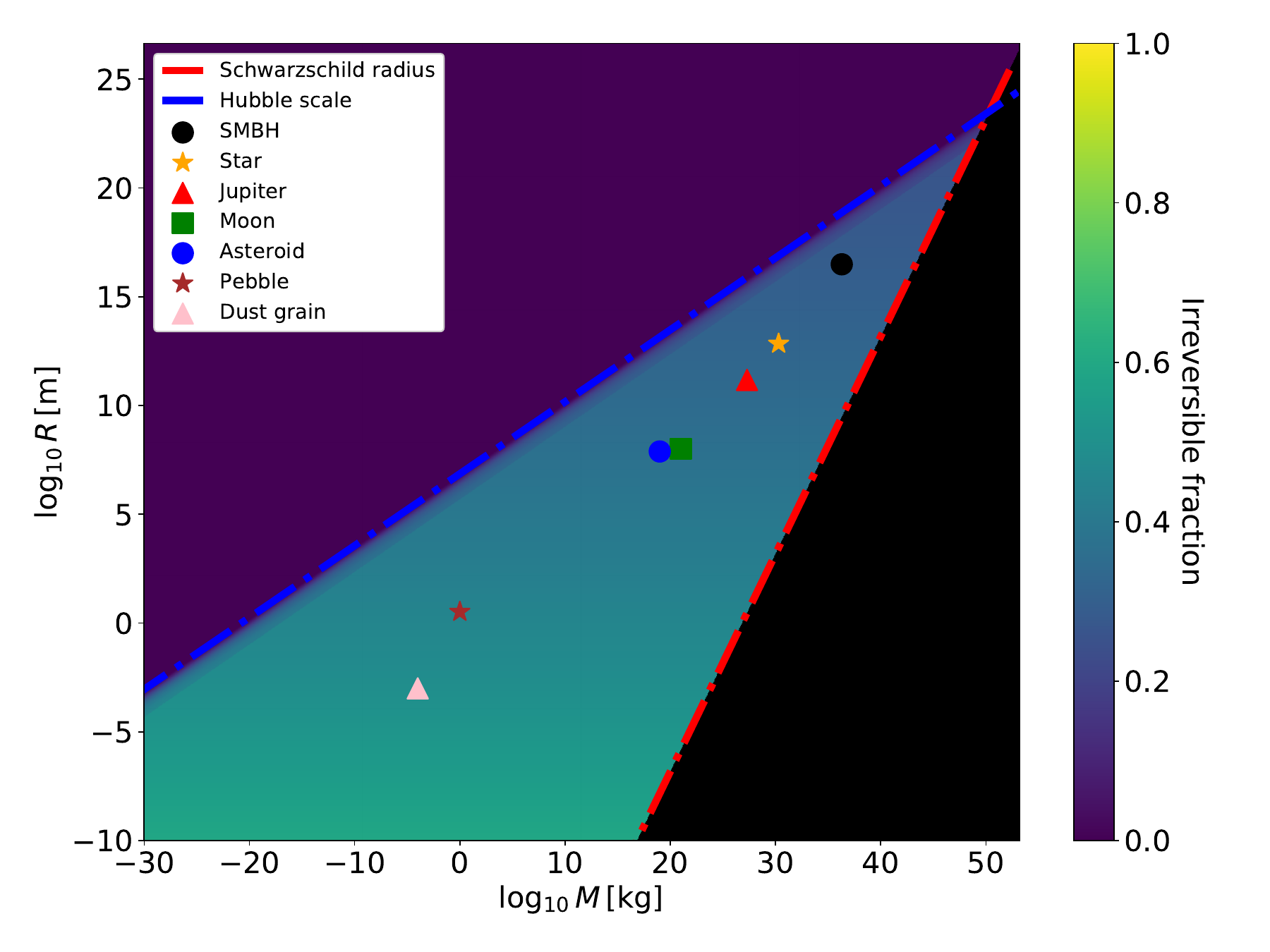} \\

\end{tabular}  
\caption{Fraction of fundamentally unpredictable triples as a function of astrophysical scale. The mass scale ranges from the electron mass to the estimated total baryonic mass in the Universe, and the length scale ranges from the size of an atom to the size of the observable Universe.
Two characteristic lines divide the region in (mostly) three large subregions. The bottom right region is excluded due to densities exceeding those of black holes (Schwarzschild radius determines the red dash-dotted line). The upper left triangular region corresponds to fundamentally predictable triples as they are unable to magnify Planck length perturbations to their own system size within a Hubble time. The analytical expression for the ``Hubble scale'' (blue dash-dotted line) is given in Eq.~\ref{eq:hubble}. In the middle region, the trend is that the fraction increases towards smaller length scales; a smaller amplification of the Planck length is required to reach the size of the triple. The locations of various astrophysical bodies are overplotted, e.g. for triple supermassive black holes (SMBH) 30\% are irreversible, and this increases to 37\% for triple asteroids and 49\% for dust grains. Here, we assumed the triples are virial ($L=0.19$), but in  Tab.~\ref{tab:triple_scale} we also provide values for the least chaotic chase ($L=0.02$). }
\label{fig:multi_scale}
\end{figure*}

\begin{figure*}
\centering
\begin{tabular}{cc}

\includegraphics[height=0.48\textwidth,width=0.64\textwidth]{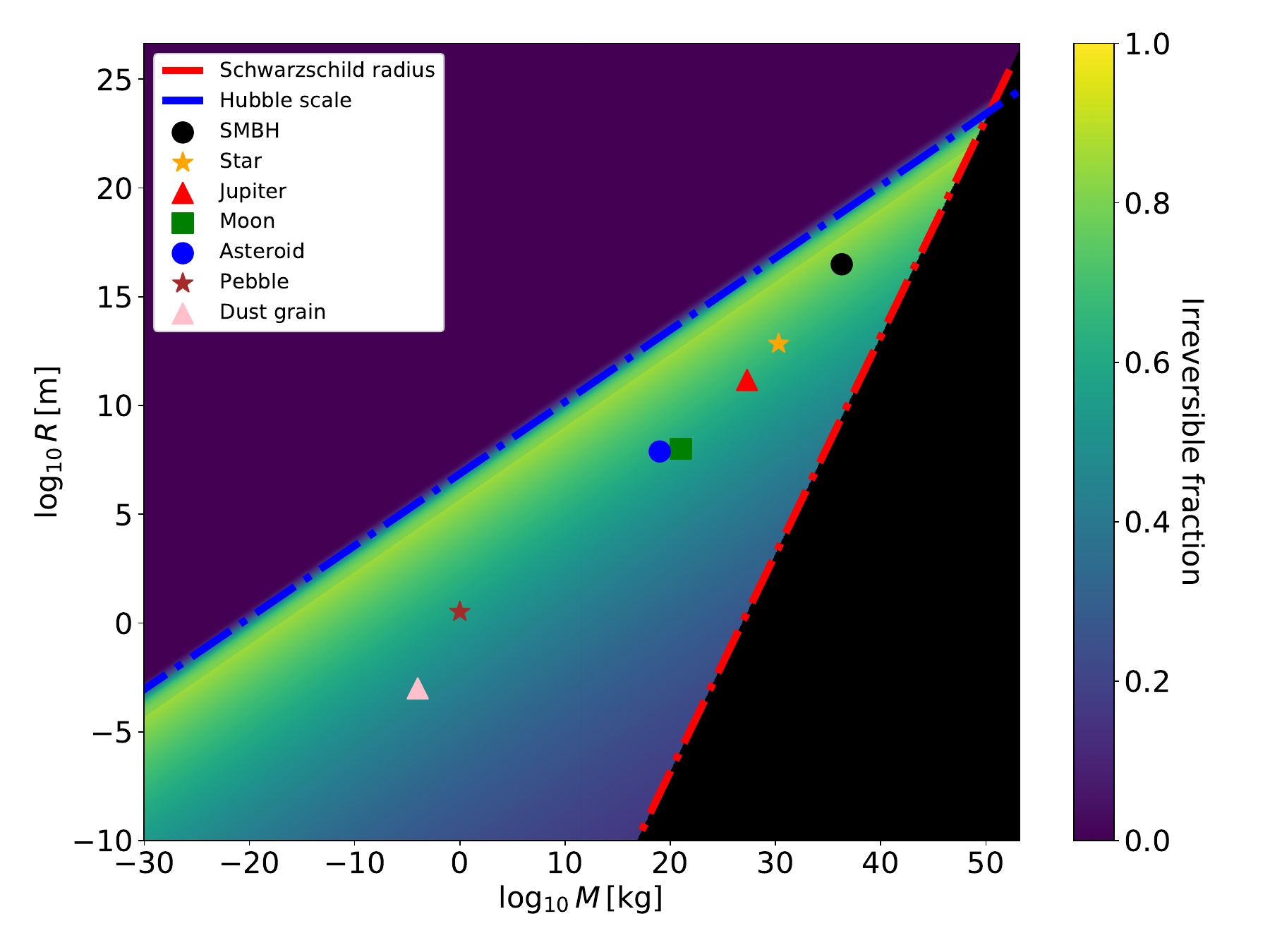} &

\end{tabular}  
\caption{Similar plot as Fig.~\ref{fig:multi_scale}, but for tidal perturbations on the triple system due to a Solar mass star at a distance of 1\,kpc. As this example shows, small tidal perturbations from other bodies in the Universe play an important role in the predictability of triple systems. Especially for the relatively loosely bound triples with the largest crossing times, i.e. near to the ``Hubble scale''. Generally, tidal perturbations tend to dominate over intrinsic quantum fluctuations. As a consequence, the fraction of irreversible triples increases; up to 77\% for triple supermassive black holes (see Tab.~\ref{tab:triple_scale}).  }
\label{fig:isolation}
\end{figure*}

It is also of interest to discuss how the fraction of fundamentally unpredictable triples depends on astrophysical scale. Although our purely Newtonian dynamical systems can be scaled up and down, here we will assume that there are physical constraints set by the Planck length and the Hubble time. 
We consider scatter plots similar to that of Fig.~\ref{fig:logA_vs_logTc_scatter}, where the Hubble time introduces a maximum cut-off in $T\,/\,T_c$, and where the Planck length introduces a minimal value of $A$ above which systems are considered to be fundamentally unpredictable. For various physical scalings of the triple system we can then count the fraction of triples in the fundamentally unpredictable region of the diagram.

The critical ``Hubble scale'' can be defined which gives the limit where a triple is just able to magnify a Planck length perturbation to its own size on a time scale of a Hubble time. The minimum amplification required is defined as the size of the triple, $R$, divided by the Planck length, $h$. This amplification is to be reached within a Hubble time, $T_H$. Using the relation between $A$ amd $T$ from Eq.~\ref{eq:fit}, we can write  

\begin{equation}
    \log_{10}A_{\rm{min}} \equiv \log_{10}\frac{R}{h} = 10^\beta \left( \frac{T_H}{T_c} \right)^\alpha,
    \label{eq:Amin}
\end{equation}

\noindent with $\alpha$ and $\beta$ the fitting parameters given in Tab.~\ref{tab:ics}. The crossing time of a virialised N-body system can be expressed as (see Sec.~\ref{sec:setup})

\begin{equation}
    T_c = \sqrt{\frac{8 R^3}{G M}},
\end{equation}

\noindent Replacing this expression into Eq.~\ref{eq:Amin} and rewriting for $M$, we obtain

\begin{equation}
    M_H \equiv 8 \times 10^{-\frac{2\beta}{\alpha}} G^{-1} T_H^{-2} R^3 \left( \log_{10}\frac{R}{h} \right)^{\frac{2}{\alpha}}.
    \label{eq:hubble}
\end{equation}

\noindent Hence, given a triple of physical size $R$, if its mass $M > M_H$, then the triple's crossing time is sufficiently small for it to be able to magnify a Planck length perturbation up to its own size. 

In Tab.~\ref{tab:triple_scale} we list various types of bodies which could be part of a triple system; from large scale supermassive black holes down to asteroids, pebbles and dust grains. For each type we give the characteristic mass and length scale, as well as their crossing time. The minimum amplification factor required to magnify a Planck length perturbation to the size of the triple itself is also given, ranging from $10^{32}-10^{51}$. In Tab.~\ref{tab:triple_scale} we also give the percentage of unpredictable triples due to quantum uncertainties, $p_h$, as a range, where the lower limit is calculated from the least chaotic ensemble ($L=0.02$) and the maximum from the most chaotic ensemble ($L=0.19$). We visualise the most chaotic case in Fig.~\ref{fig:multi_scale}, where we also overplot the locations of the various astrophysical bodies. 
The general trend is that the fraction of fundamentally unpredictable triples increases for more compact triples.
Smaller systems require a smaller amplification of the Planck length in order to become unpredictable. 

Hence, by increasing the angular momentum of supermassive black hole triples, the fraction of unpredictable systems is enhanced from $5\%$ to $30\%$, and by decreasing the scale of the triple down to triple dust grains, the percentage is further enhanced to about $50\%$.
Fundamentally unpredictable triples are general, occurring over a large range of astrophysical scales owing to the exponential nature of chaos.   

\subsection{Small perturbations from the rest of the Universe}\label{sec:isolation}

In Paper 1 and in this current study, we focused on perturbations due to intrinsic quantum uncertainties. We assumed that these form the smallest physical perturbations in nature, and demonstrated that they still play a role in the predictability of triple systems (and therefore in larger-$N$ systems too). However, there are other sources of perturbations in the Universe, which can be much larger in magnitude. For example, we can compare the evolution of an isolated triple system to one with a fourth-body at some large distance from the triple. This external body will induce a tidal effect onto the triple, causing each of the triple components to experience a slightly different tidal acceleration. In the center of mass frame of the triple this will manifest as slight perturbations in the orbits of the bodies (compared to the isolated triple case). The magnitude of these tidal seed perturbations depends on the distance of the fourth body, and we ask at what distance does the tidal perturbation become of the same order as the Planck length? The tidal acceleration is estimated as

\begin{equation}
    \delta a \sim \frac{G m_t R}{\left(\gamma R\right)^3} = \frac{G m_t}{\gamma^3 R^2},
\end{equation}

\noindent with $G$ the gravitational constant, $m_t$ the mass of the tidal perturber, and $R$ the size of the triple. The factor $\gamma$ gives the distance of the fourth-body in units of $R$. The accelerations within the triple are of order $a \sim \frac{GM}{R^2}$, with $M$ the mass of the triple. We can estimate the seed perturbation in the orbit according to $\frac{\delta r}{R} \sim \frac{\delta a}{a}$, allowing us to write

\begin{equation}
    \delta r \sim R \frac{\delta a}{a} = \frac{R}{\gamma^3} \frac{m_t}{M}.
    \label{eq:dr}
\end{equation}

\noindent Setting $\delta r$ equal to the Planck length, $h$, we obtain the characteristic separation of 

\begin{equation}
    \gamma_h = \left( \frac{R}{h} \frac{m_t}{M} \right)^{\frac{1}{3}}.
\end{equation}

\noindent We compare this separation to the size of the observable Universe, $R_U = \gamma_{U} R$, so that we finally obtain

\begin{equation}
    \frac{\gamma_h}{\gamma_U} = \left( \frac{R^4}{h R_U^3} \frac{m_t}{M} \right)^{\frac{1}{3}}.
\end{equation}

\noindent Now we will assume that the fourth body is of the same type as the triple components, i.e. $m_t \sim M$. Furthermore, by setting $\gamma_h = \gamma_U$, we obtain the following characteristic length scale:

\begin{equation}
    R_h \equiv h^\frac{1}{4} R_U^\frac{3}{4} \approx 1\,\rm{au}.
\end{equation}

\noindent Triple systems larger than this surprisingly small length scale, will be susceptible to tidal perturbations of order the Planck length due to bodies beyond the cosmological horizon. For more compact triples, such distant tidal perturbations will be negligible compared to the intrinsic quantum uncertainty. 

In reality, there will be many tidal perturbers much closer to home. For example, for our standard test case supermassive black hole triple, there are billions of perturbing stars in the host galaxy. Using Eq.~\ref{eq:dr}, with $M=10^6\,\rm{M_\odot}$, $R = 1\,\rm{pc}$, $m_t = 1\,\rm{M_\odot}$, and $\gamma = 10^3$, we estimate a seed perturbation of $\delta r = 30\,\rm{m}$ in the orbits of the black holes. Magnifying this to the size of the triple requires an amplification factor of only $R / \delta r \sim 10^{15}$. 
Similar to the previous subsection, we can count the fraction of triple systems that reach an amplification factor larger than $10^{15}$ within a Hubble time, which ranges from $49-77\%$ depending on $L$. For smaller scale triples, the fraction of unpredictable triples due to tidal perturbations tends to decrease, e.g. down to $25-62\%$ for asteroids and $12-48\%$ for dust grains (see Tab.~\ref{tab:triple_scale}). In Fig.~\ref{fig:isolation}, we visualise these percentages and we observe that tidal perturbations mostly affect the loosely bound triples, i.e. those with the largest crossing times near the Hubble scale. 
Generally, external tidal perturbations from other bodies in the Universe tend to dominate over intrinsic quantum uncertainties. 

\subsection{Future work}

The contradiction between the naive expectation that lower-$L$ systems would be maximally chaotic, and our new numerical results motivates further investigation into the origin of chaos in triple systems (as well as larger N-body systems \citep[e.g.][]{SPZ2022, 2022arXiv220903347P}. Although it is well known that higher angular momentum triples tend to live longer on average, here we find that they can also have larger maximum Lyapunov exponents (shorter Lyapunov times). The driver of chaos is therefore not (solely) close encounters in radial orbits, but rather the prolonged and non-linear interaction among all three bodies in a democratic configuration. However, we find that even during such democratic resonances the rate of divergence can vary. Based on our numerical results, we speculate that high angular momentum triples might statistically  have a longer cumulative resonant interaction time and/or shorter excursion phases of a single body. This requires a closer inspection of the dependence of the instantaneous Lyapunov exponent on the specifics of the orbital configuration. 

\section*{Acknowledgements}

We acknowledge the valuable feedback received on this project during the Chaotic Rendez-Vous meeting in Edinburgh in 2023 organised by Anna Lisa Varri. The calculations were performed using the LGMII (NWO grant \#621.016.701). Part of this publication is funded by the Nederlandse Onderzoekschool Voor Astronomie (NOVA). 

\section*{Data Availability}

The data underlying this article will be shared on reasonable request to the corresponding author.



\bibliographystyle{mnras}
\bibliography{agekyan} 





\bsp	
\label{lastpage}
\end{document}